\documentclass[twocolumn]{aastex63}

\usepackage[super]{nth}
\usepackage{savesym}
\savesymbol{tablenum}
\usepackage[T1]{fontenc}
\usepackage{inputenc}
\usepackage[range-units = brackets,tophrase={-},seperr,repeatunits=false]{siunitx}
\restoresymbol{SIX}{tablenum}

\graphicspath{{./}{figures/}}

\received{\today}

\submitjournal{ApJ}

\shorttitle{Molecular Gas in the NGC~5044 Group}
\shortauthors{Schellenberger et al.}

\begin{document}

\title{Atacama Compact Array Measurements of the Molecular Mass in the NGC~5044 Cooling Flow Group}
\correspondingauthor{Gerrit Schellenberger}
\email{gerrit.schellenberger@cfa.harvard.edu}

\author[0000-0002-4962-0740]{Gerrit Schellenberger}
\affiliation{Center for Astrophysics $|$ Harvard \& Smithsonian, 60 Garden St., Cambridge, MA 02138, USA}

\author{Laurence P. David}
\affiliation{Center for Astrophysics $|$ Harvard \& Smithsonian, 60 Garden St., Cambridge, MA 02138, USA}

\author{Jan Vrtilek}
\affiliation{Center for Astrophysics $|$ Harvard \& Smithsonian, 60 Garden St., Cambridge, MA 02138, USA}

\author[0000-0002-5671-6900]{Ewan O'Sullivan}
\affiliation{Center for Astrophysics $|$ Harvard \& Smithsonian, 60 Garden St., Cambridge, MA 02138, USA}

\author{Jeremy Lim}
\affiliation{Department of Physics, University of Hong Kong, Pokfulam Road, Hong Kong}

\author[0000-0002-9478-1682]{William Forman}
\affiliation{Center for Astrophysics $|$ Harvard \& Smithsonian, 60 Garden St., Cambridge, MA 02138, USA}

\author[0000-0001-5880-0703]{Ming Sun}
\affiliation{Department of Physics, University of Alabama in Huntsville, Huntsville, AL 35899, USA}

\author[0000-0003-2658-7893]{Francoise Combes}
\affiliation{Observatoire de Paris, LERMA, CNRS, 61 Avenue de l'Observatoire, 75014 Paris, France}

\author{Philippe Salome}
\affiliation{Observatoire de Paris, LERMA, CNRS, 61 Avenue de l'Observatoire, 75014 Paris, France}

\author{Christine Jones}
\affiliation{Center for Astrophysics $|$ Harvard \& Smithsonian, 60 Garden St., Cambridge, MA 02138, USA}

\author[0000-0002-1634-9886]{Simona Giacintucci}
\affiliation{Naval Research Laboratory, 4555 Overlook Avenue SW, Code 7213, Washington, DC 20375, USA}

\author[0000-0002-3398-6916]{Alastair Edge}
\affiliation{Institute for Computational Cosmology, Department of Physics, Durham University, South Road, Durham DH1 3LE}

\author[0000-0002-9112-0184]{Fabio Gastaldello}
\affiliation{INAF - IASF-Milano, Via E. Bassini 15, 20133 Milano, Italy}

\author[0000-0002-8341-342X]{Pasquale Temi}
\affiliation{Astrophysics Branch, NASA/Ames Research Center, MS 245-6, Moffett Field, CA 94035}

\author[0000-0001-9807-8479]{Fabrizio Brighenti}
\affiliation{Dipartimento di Astronomia, Universita di Bologna, via Ranzani 1, Bologna 40127, Italy}

\author[0000-0002-8900-0298]{Sandro Bardelli}
\affiliation{INAF - Astrophysics and Space Science Observatory Bologna, via Gobetti 93/3, 40129 Bologna, Italy}

\begin{abstract}
The fate of cooling gas in the centers of galaxy clusters and groups is still not well understood, as is also the case for the complex processes of triggering star formation in central dominant galaxies (CDGs), re-heating of cooled gas by AGN, and the triggering/``feeding'' of supermassive black hole outbursts.  
We present CO observations of the early type galaxy NGC~5044, which resides at the center of an X-ray bright group with a moderate cooling flow. For our analysis we combine CO(2-1) data from the 7m antennae of the Atacama Compact Array (ACA), and the ACA total power array (TP). We demonstrate, using the 7m array data, that we can recover the total flux inferred from IRAM 30m single dish observations, which corresponds to a total molecular mass of about $\SI{4e7}{M_\odot}$. 
Most of the recovered flux is blueshifted with respect to the galaxy rest frame and is extended on kpc-scales, suggesting low filling factor dispersed clouds. 
We find 8 concentrations of molecular gas out to a radius of $\SI{10}{arcsec}$ ($\SI{1.5}{kpc}$), which we identify with giant molecular clouds. The total molecular gas mass is more centrally concentrated than the X-ray emitting gas, but extended in the north-east/south-west direction beyond the IRAM 30m beam. We also compare the spatial extent of the molecular gas to the H$\alpha$ emission: The CO emission coincides with the very bright H$\alpha$ region in the center. We do not detect CO emission in the fainter H$\alpha$ regions. Furthermore, we find two CO absorption features spatially located at the center of the galaxy, within $\SI{5}{pc}$ projected distance of the AGN, infalling at 255 and $\SI{265}{km\,s^{-1}}$ relative to the AGN. This indicates that the two giant molecular clouds seen in absorption are most likely within the sphere of influence of the supermassive black hole.
\end{abstract}


\section{Introduction} \label{sec:intro}
The increasing density of hot X-ray emitting gas towards the centers of galaxy clusters and groups predicts short cooling times resulting in large amounts of cold gas (e.g., \citealp{Fabian1994-oe,Salome2003-ge,David2014-jn}).
Also, simulations show that the densest parts of the H$\alpha$ filaments can cool to molecular clouds with masses of several $\SI{1e7}{M_\odot}$ (\citealp{Gaspari2017-ev}).
Active galactic nuclei (AGN) have been found to be a significant re-heating source, and may resolve the conflict between larger apparent cooling rates and minimal star formation in X-ray luminous groups and clusters.
Cold molecular gas has been found to form in the central dominant galaxies (CDGs) of several systems (e.g., \citealp{Edge2001-mi,Salome2003-ge}, and more recently \citealp{Tremblay2016-xg,Vantyghem2017-ym,OSullivan2018-yk,Rose2019-tu}). 
In these studies single dish CO measurements have turned out to be essential to reveal the total molecular gas resulting from cooling in the centers of groups and clusters (\citealp{McNamara2007-xa,Russell2014-su,McNamara2016-ux,Russell2016-ek,Russell2017-xk,Russell2017-zd}). However, in order to make a comparison to feedback models, a more precise location of these cold gas reservoirs is needed: Are they associated in small or large clouds, or fully diffuse? High resolution interferometry measurements (e.g., with the Atacama Large Millimeter/submillimeter Array, ALMA) have found small (wrt the total CO flux) clouds of CO (\citealp{David2014-jn,Temi2018-rr,Olivares2019-rs}). However, the integrated mass of these clouds is inconsistent with the single dish measurements, indicating that the large baselines of ALMA are unable to capture a large fraction of the molecular gas (\citealp{Olivares2019-rs}). ALMA alone cannot provide a comprehensive picture of the cold gas in clusters and groups. The Atacama Compact Array (ACA) with its shorter baselines closes the gap with single dish measurements. The ACA is composed of 12 7m antennae (7m array) used in interferometry mode and 4 12m antennae used in total power mode (TP array). The configuration of the ACA ensures superb coverage of short baselines, and a spatial resolution close to  $\SI{5}{\arcsec}$.

NGC~5044 is one of the X-ray brightest galaxy groups in the sky. The wealth of multifrequency data available for this object makes it an ideal candidate to study the correlations of gas properties over a broad range of temperatures. H$\alpha$ filaments, [CII] line and CO emission show that some gas must be cooling out of the hot phase (\citealp{David2014-jn,Werner2014-vw}). Prior studies using exceptionally deep, high-resolution X-ray data have shown that the hot gas within the central region has been perturbed by at least three cycles of AGN outbursts (\citealp{David2009-hn,Giacintucci2011-ru,David2017-ig}).
\cite{David2014-jn} have spatially resolved clumps of molecular gas from CO 2-1 line emission in the central early type galaxy with ALMA. 
Also in NGC~5044 most of the emission that has been detected in single dish observations with the IRAM 30m telescope was missed by the early ALMA measurements, despite ALMA having lower noise levels and thus higher sensitivity. However, the ALMA observation misses the shortest baselines needed to recover the full flux.
In order to understand the cooling processes in the centers of clusters and groups, one needs to have a full census of the giant molecular clouds, including their precise position in spatial and velocity space, to enable a useful comparison with competing models for the feedback process. This requires observations with interferometers in multiple array configurations to cover the different sizes and scales of molecular clouds.
H$\alpha$ and [CII] $\lambda \SI{158}{\mu m}$ observations (\citealp{Werner2014-vw}) show that in the case of NGC~5044 the filaments of cooling gas extend out to 8\,kpc, while multiphase gas in the X-rays has been detected out to 15\,kpc (\citealp{David2017-ig}). Cold gas CO measurements confirm the existence of Giant Molecular Clouds (\citealp{David2014-jn}) in the central 2\,kpc, with sizes below 800\,pc. However, single dish measurements show that most of the gas must be on larger scales.

Our early (Cycle 0) ALMA observation of NGC~5044 has also been analyzed by \cite{Temi2018-rr}, who found results consistent with \cite{David2014-jn} for the most significant clouds in terms of cloud location and size. Almost all central clouds and the ones located north-west of the AGN are confirmed, while 9 clouds, mostly in the east, are unconfirmed and 2 new ones are detected. The physical properties of the confirmed ones are in good agreement with \cite{David2014-jn}.
To extend these measurements in terms of covered spatial scales, we present new ACA observations of the galaxy group NGC~5044.

In section \ref{ch:data_reduction} we describe the ACA data products and processing. Section \ref{ch:results} presents our main results, including a comparison of the 7m array spectrum with the IRAM 30m spectrum, estimates of the total molecular mass in the system and on which scales the gas is distributed, the detection of clouds and a comparison to the \cite{David2014-jn} findings, and the two absorption features in the spectrum. Section \ref{ch:discussion} gives a comparison of the detected molecular clouds with the ones detected in other systems and simulations, and discusses the virial equilibrium in these clouds. We also constrain the physical distance of the clouds seen in absorption, and the implication for AGN feedback. Our findings are summarized in section \ref{ch:summary}.

In this paper we assume a flat $\Lambda$CDM cosmology with $H_0 = \SI{70}{km\,s^{-1}\,Mpc^{-1}}$, and $\Omega_\mathrm{m}=0.3$.
We adopt a systemic velocity of  $\SI{2757}{km\,s^{-1}}$ (heliocentric) for NGC~5044 and a luminosity distance of $\SI{31.2}{Mpc}$ (\citealp{Tonry2001-wi}), which gives a physical scale in the rest frame of NGC~5044 of $\SI{1}{\arcsec}=\SI{150}{pc}$. Uncertainties are given at the $1\sigma$ level throughout the paper.

\section{ALMA Data Reduction}\label{ch:data_reduction}
\begin{figure*}
\center{\includegraphics[width=.99\linewidth]{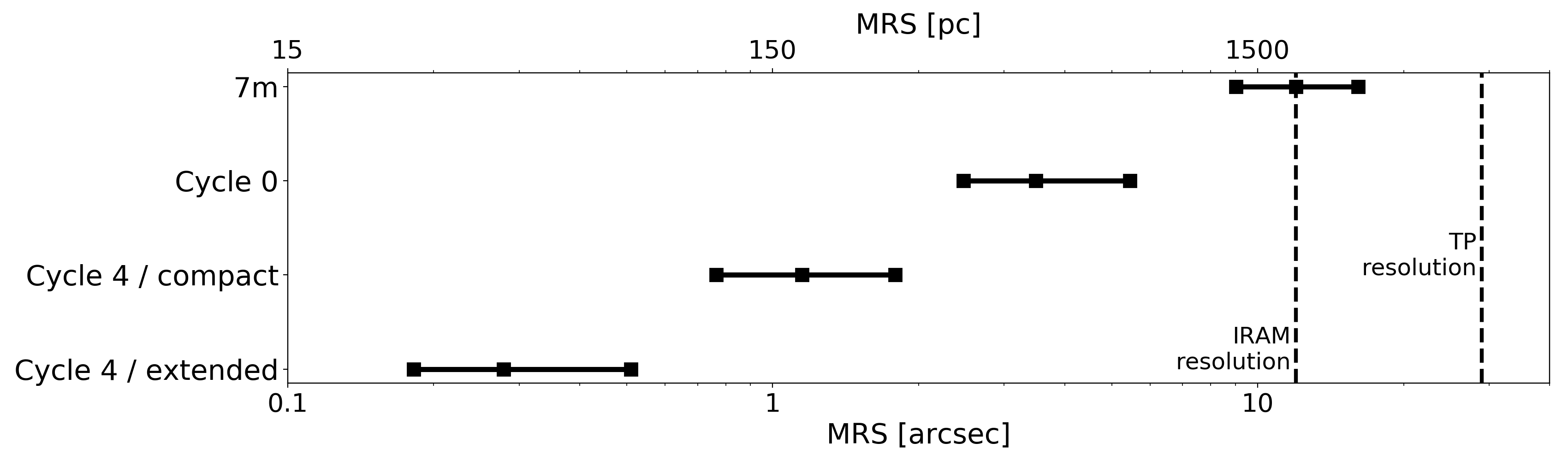}}
\caption{The three points along each horizontal line are maximum recoverable scales (MRS) that correspond to the 25\%, 50\%, and 75\% percentiles in the cumulative baseline distribution for each array configuration. Angular scales are shown at the bottom and physical scales in the rest frame of NGC~5044 at the top. Nearly complete coverage of all angular scales is provided by the different ALMA/ACA array configurations. }
\label{fig:arrays}
\end{figure*}

NGC~5044 was observed in cycle 4 with two different ALMA configurations (compact and extended) and with the ACA (7m array plus antennae for total power measurements). 
An observing log of all ALMA and ACA CO(2-1) observations of NGC~5044, including the cycle 0 observation and the recent total power (TP) ALMA single dish observation, is given in Table \ref{tab:obslognew}. The last row in Tab. \ref{tab:obslognew} indicates the usage of each dataset in this paper. 
Note that the detailed analysis of the cycle 4 12m array data will be presented in a subsequent paper. However, we list here the cycle 4 12m array observations for completeness. Only the highest resolution cycle 4 observations are used for the absorption study in section \ref{ch:absorption}.
Details about the ALMA cycle 0 observation can be found in \cite{David2014-jn}. 
The two ALMA observations taken in cycle 4 were mosaics of 11 pointings with a Nyquist sampling covering a region of approximately $\SI{60}{\arcsec}$ by $\SI{60}{\arcsec}$ centered on NGC~5044. 
The 7m observation was a single pointing centered on NGC~5044 and covering a region of about $40\,\arcsec$ diameter. 
The TP observation covers a region of about $\SI{55}{\arcsec}\times \SI{55}{\arcsec}$ scanned in raster mode. 
Fig. \ref{fig:arrays} shows the range of angular and physical scales sampled by the ACA and ALMA for the different array configurations. 
As can be seen in this figure, the combination of all ALMA and ACA observations provides a nearly complete census of the total molecular mass and the detection of individual molecular clouds in NGC~5044 on scales up to the TP field of view.
We use the 7m array data for comparisons between the ALMA and IRAM measurements, showing that the total single dish flux is recovered in the interferometer observations (e.g., Figs. \ref{fig:12asec_spec}, \ref{fig:mass_mrs}). We also use the 7m array image cubes for the cloud detection, to have slightly higher spatial resolution. 
We combine the 7m array data with the Total Power array (7m + TP $=$ ACA) to measure the total CO emission at larger radii ($>12\si{arcsec}$), where the 7m array sensitivity drops. In the following we describe the processing steps beyond the standard pipeline.

\subsection{7m array}
All ACA and ALMA observations were configured to provide four spectral windows within Band 6, each with a bandwidth of $\SI{185}{MHz}$.  
One of the spectral windows was centered on the CO(2-1) line at \SI{228.397}{GHz}.  The central frequencies for the other three spectral windows were  $\SI{226.418}{GHz}$, $\SI{242.314}{GHz}$, and $\SI{244.244}{GHz}$.
The correlator configuration was similar for all cycle 4 observations. 

The 7m data were calibrated with CASA version 4.7.2 (\citealp{McMullin2007-ed}) using the ALMA pipeline scripts, but CASA version 5.4.0 was used for the subsequent imaging. 
Several phase self-calibration steps improved the noise level of the continuum image by a factor of 2.7, and the final solutions were applied to the spectral window containing the CO(2-1) emission. Continuum subtraction was performed with the CASA tool \verb|uvcontsub| using the emission in the line-free regions. The CASA tool \verb|tclean| was used to generate data cubes from the continuum subtracted measurement set.  Unless otherwise noted, we used a Briggs weighting with a robust parameter of 1, and the \verb|multiscale| deconvolver (\citealp{Cornwell2008-no}, with scales set to 0 and 3). 
For the iterative \textit{clean} algorithm we applied the \textit{auto-multithresh} masking algorithm with the \textit{noisethreshold}$=4.5$, and \textit{sidelobethreshold}$=1.25$, to clean until a threshold of $0.3\sigma$ is reached.
The synthesized beams and the root mean square (rms) sensitivities in $\SI{10}{km\,s^{-1}}$ velocity channels are shown in Table \ref{tab:obslognew} for all data sets, although we focus here on the ACA data. 

\subsection{Total Power array}
The TP array data were recorded in nine schedule blocks between October and December 2018, out of which six were recorded using four antennae, two blocks with three antennae, and one block with two antennae. The science spectral window contains 2048 channels over a $\SI{2}{GHz}$ bandwidth. We used CASA version 5.4.0 to apply the pipeline calibration and flagging of bad data. 
On-off single dish calibration was used to measure $T_\mathrm{sys}$, and NGC~5044 was scanned in raster mode.

Imaging was achieved through the CASA task \verb|sdimaging| with a spheroidal gridding function and $\SI{10}{km\,s^{-1}}$ channel width.
We ran the pipeline task \verb|hsd_baseline| to subtract the baseline inferred through position switching.
The baseline is not flat due to spatial and spectral variability in the response. After creating an image cube we used \verb|imcontsub| on the line free channels to fully subtract continuum emission.

In order to combine the 7m and TP array image cubes we regridded the TP image cube to match the 7m dimensions, and divided by the 7m primary beam (PB) to have a common PB attenuation. After combining the image cubes with the CASA task \verb|feather| we corrected the resulting image cube for the 7m PB.

\begin{deluxetable*}{cccccc}
 \tablecaption{ALMA Observing Log}
 \tablehead{
 \colhead{} &
 \colhead{12m} &
 \colhead{12m} &
 \colhead{7m} & 
 \colhead{12m} &
 \colhead{Total Power} \\
 \colhead{} &
 \colhead{Cycle 0} &
 \colhead{Cycle 4 extended} &
 \colhead{Cycle 4} & 
 \colhead{Cycle 4 compact} &
 \colhead{} 
 }
\startdata
Date & 2012 Jan 13  & 2017 Aug 3 & 2016 July & 2016 Nov 15 & 2018 Oct -- Dec\\
Project Code & 2011.0.00735.S & 2016.1.00533.S& 2016.2.00134.S & 2016.1.00533.S & 2017.1.00784.S\\
Time (min) & 29 & 44 & 306 & 48 & 614 \\
Bandpass & 3C273 & J1337-1257 & J1256-0547 & J1337-1257 & -\\
Phase  & J1337-129 & J1258-1800 & J1337-1257 & J1305-1033 & -\\
Amplitude & Titan & Titan & Titan & Titan & - \\ 
\# Ant & 18 & 46 & 11 & 42 & 2--4 \\
rms ($\si{mJy\,bm^{-1}}$) & 1.4 & 0.49 & 1.8 & 0.47 & 12 \\
Beam size($\arcsec \times \arcsec$) & 2.1 x 1.1 & 0.13 x 0.11 & 6.9 x 5.1 & 0.54 x 0.47 & 28.4 x 28.4\\
Beam size(pc$\times$pc) & 315 x 165 & 20 x 17 & 1050 x 765 & 81 x 70 & 4300 x 4300\\
$\Delta v$ ($\si{km\,s^{-1}}$)& 0.64 & 1.48 & 0.59 & 1.48 & 1.48\\
Used here & flux comparison & absorption & flux \& new clouds & not used & together with 7m
\enddata
\tablecomments{Observation date, ALMA project code, on target science integration time, calibrators, number of antennae, rms sensitivity for channel width $\SI{10}{km\,s^{-1}}$, restoring beam, velocity resolution, and application in this paper. For the total power observations three calibrators for the water vapor radiometry were used: J1245-1616, J1337-1257, J1305-1033. \label{tab:obslognew}}
\end{deluxetable*}

\section{ACA Results}\label{ch:results}
\begin{figure*}
    \centering
    \includegraphics[width=0.8\linewidth]{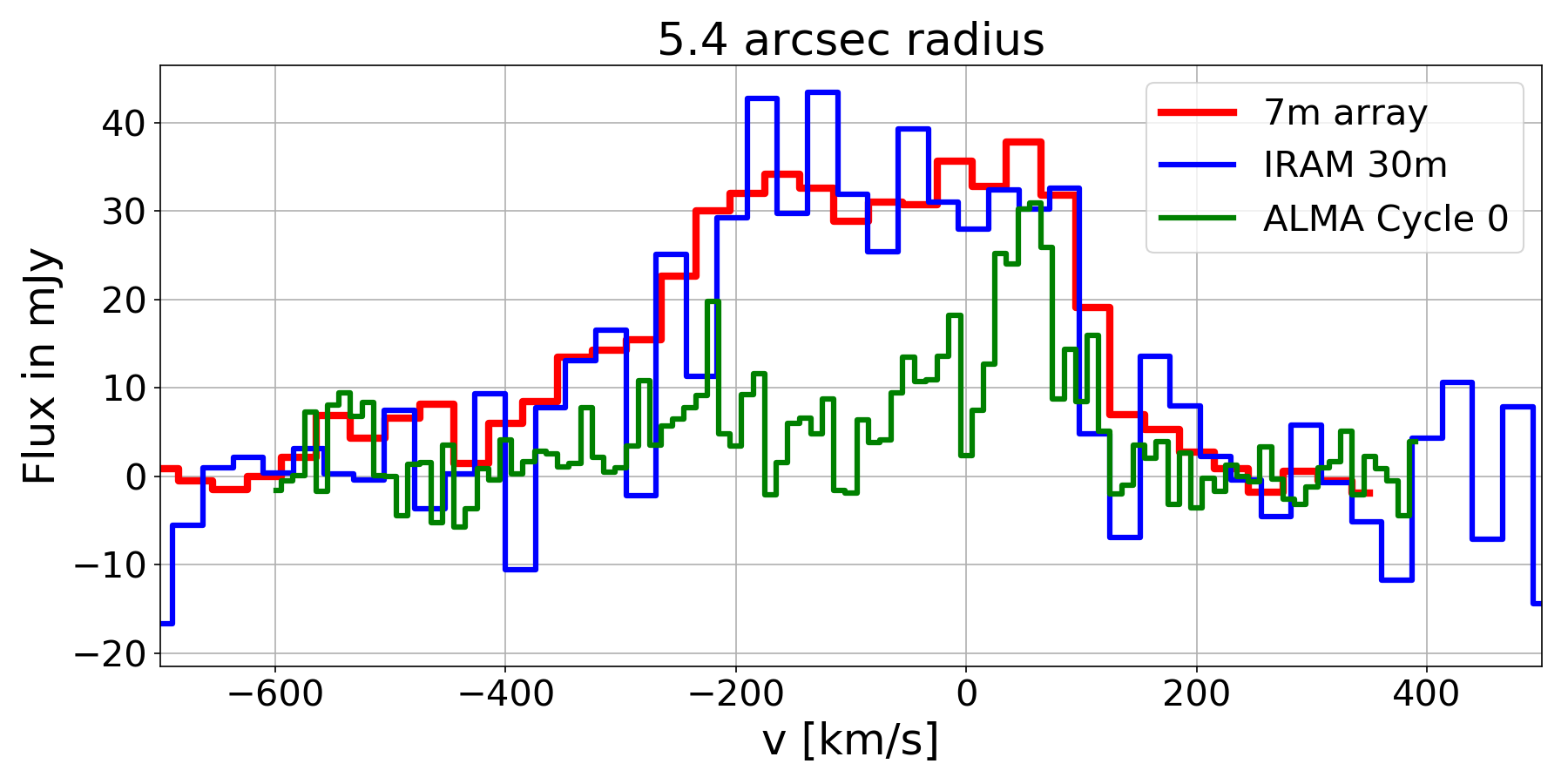}
    \caption{Comparison of IRAM 30m (blue), 7m array (red), and ALMA 12m Cycle 0 (green) CO(2-1) spectra. All ALMA spectra are extracted within a $5.4^{\prime\prime}$ radius aperture (i.e., the IRAM 30m PB). The 7m observation recovers all the flux measured by the IRAM single dish telescope.}
    \label{fig:12asec_spec}
\end{figure*}

\begin{figure}
    \centering
    \includegraphics[width=0.99\linewidth]{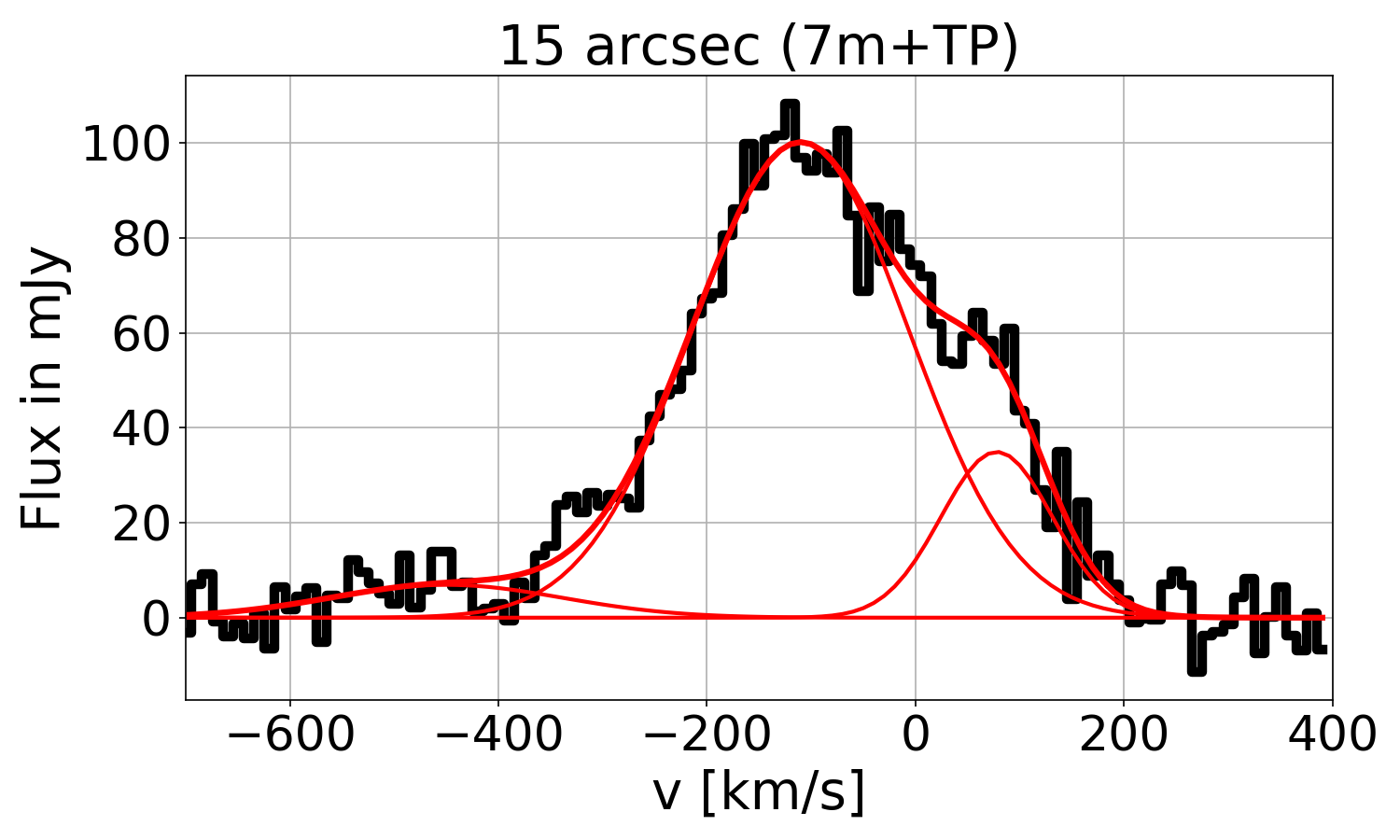}
    \caption{ALMA 7m plus Total Power Array spectrum (combined through feathering) within $\SI{15}{\arcsec}$ radius aperture (black). The three Gaussian components are shown in red, and the values are given in Tab. \ref{tab:fit30arcsec}. The spectrum is well characterized by the sum of the three broad Gaussians.}
    \label{fig:30asec_spec}
\end{figure}

The primary objective of our ACA 7m array observation was to recover the CO(2-1) emission that was resolved out in our earlier ALMA observation, but detected in the single dish IRAM 30\,m observation.
A comparison between the IRAM 30m, ALMA cycle 0, and 7m spectra is shown in Fig. \ref{fig:12asec_spec}.
For this purpose, we also reprocessed the cycle 0 data with the latest calibration pipeline, and performed a phase self calibration, which improved the noise level by a factor of 1.4. 
The 7m array and ALMA spectra were extracted from circular apertures with a $\SI{6}{\arcsec}$ radius (i.e., the IRAM 30m PB). 
This comparison shows that essentially all of the CO(2-1) emission is recovered in the 7m observation, indicating that the bulk of the CO(2-1) emission in NGC~5044 arises from diffuse molecular structures with spatial scales of above 1\,kpc.  
The 7m observation also detects significant CO(2-1) emission beyond the IRAM 30m PB (see Figs. \ref{fig:30asec_spec} and \ref{fig:ACA_specbins}), with a total extent of approximately $\SI{30}{\arcsec}$.  
Figure \ref{fig:30asec_spec} shows that the ACA  CO(2-1) spectrum within $15^{\prime\prime}$ radius is well represented by the sum of three broad Gaussians (also see Table \ref{tab:fit30arcsec}). 
Note that the signal to noise ratios (SNR) in Table \ref{tab:fit30arcsec} reflect the integrated signal within the FWHM and the noise in the combined bins of the FWHM of each Gaussian component.
The redshifted complex was the most massive structure detected in the ALMA cycle 0 data and it was identified as GMC 18 by \cite{David2014-jn}.
While the blueshifted molecular complex centered at $\SI{-110}{km\,s^{-1}}$ is the most massive of the three complexes, it was almost entirely resolved out in the ALMA cycle 0 observation.
The more sensitive ACA observation also detects a blueshifted molecular complex at approximately $\SI{-450}{km\,s^{-1}}$ which almost exceeds the adiabatic sound speed of the hot gas ($c_a = \SI{460}{km\,s^{-1}}$). 
We use a likelihood-ratio analysis to verify the significance of this third component at high (negative) velocity. We find that the null hypothesis (the spectrum is described by only two Gaussians) is rejected by $p_0 < \num{e-4}$, when compared to a three component fit. This is even more significant if we use a smaller aperture (10 instead of $\SI{15}{arcsec}$.)

\begin{deluxetable}{cccccc}
 \tablecaption{Three Gaussian Fit to ACA Spectrum}
 \tablehead{
 \colhead{v$_0$} &
 \colhead{FWHM} &
 \colhead{Amp} &
 \colhead{S$_{CO}$ $\sigma$} &
 \colhead{M$_{mol}$} & 
 \colhead{SNR}
 \\
 \colhead{(km\,s$^{-1}$)} &
 \colhead{(km\,s$^{-1}$)} &
 \colhead{(mJy)} &
 \colhead{(Jy\,km\,s$^{-1}$)} &
 \colhead{($10^7$~M$_{\odot}$)}  &
 \colhead{}
 }
\startdata
$\num{-450(58)}$ & $\num{263(132)}$ & \phn \phn $\num{7(2)}$ & \phn $\num{1.8(9)}$ & $\num{0.6(3)}$ & 5 \\
$\num{-110(5)}$\phn  & $\num{244(19)}$\phn  & $\num{100(2)}$ & $\num{26.0(19)}$ & $\num{8.2(6)}$ & 67\\
\phm{$-$}$\num{78(8)}$ & $\num{126(18)}$\phn  & \phn $\num{35(6)}$ & \phn $\num{4.7(13)}$ & $\num{1.5(4)}$ & 16 \\
\enddata
\tablecomments{Velocity centroid, full width half maximum, amplitude, integrated
CO(2-1) flux density, and molecular mass for the three Gaussian fit to the 
ACA (7m+TP) spectrum extracted within a  $\SI{15}{\arcsec}$ radius aperture (as shown in Fig. \ref{fig:30asec_spec}). The total molecular mass is $\SI{10.3(2)e7}{M_\odot}$.\label{tab:fit30arcsec}}
\end{deluxetable}

\subsection{Moments of the CO(2-1) Emission}
In this section we investigate the large scale properties of the molecular emission in NGC~5044 through the CO moment maps, before we compare the various total molecular mass estimates in Section \ref{ch:total_mol_mass}, and study the emission in the individual channel maps in Section \ref{ch:channelmaps}. 

\begin{figure}
\includegraphics[width=0.99\linewidth]{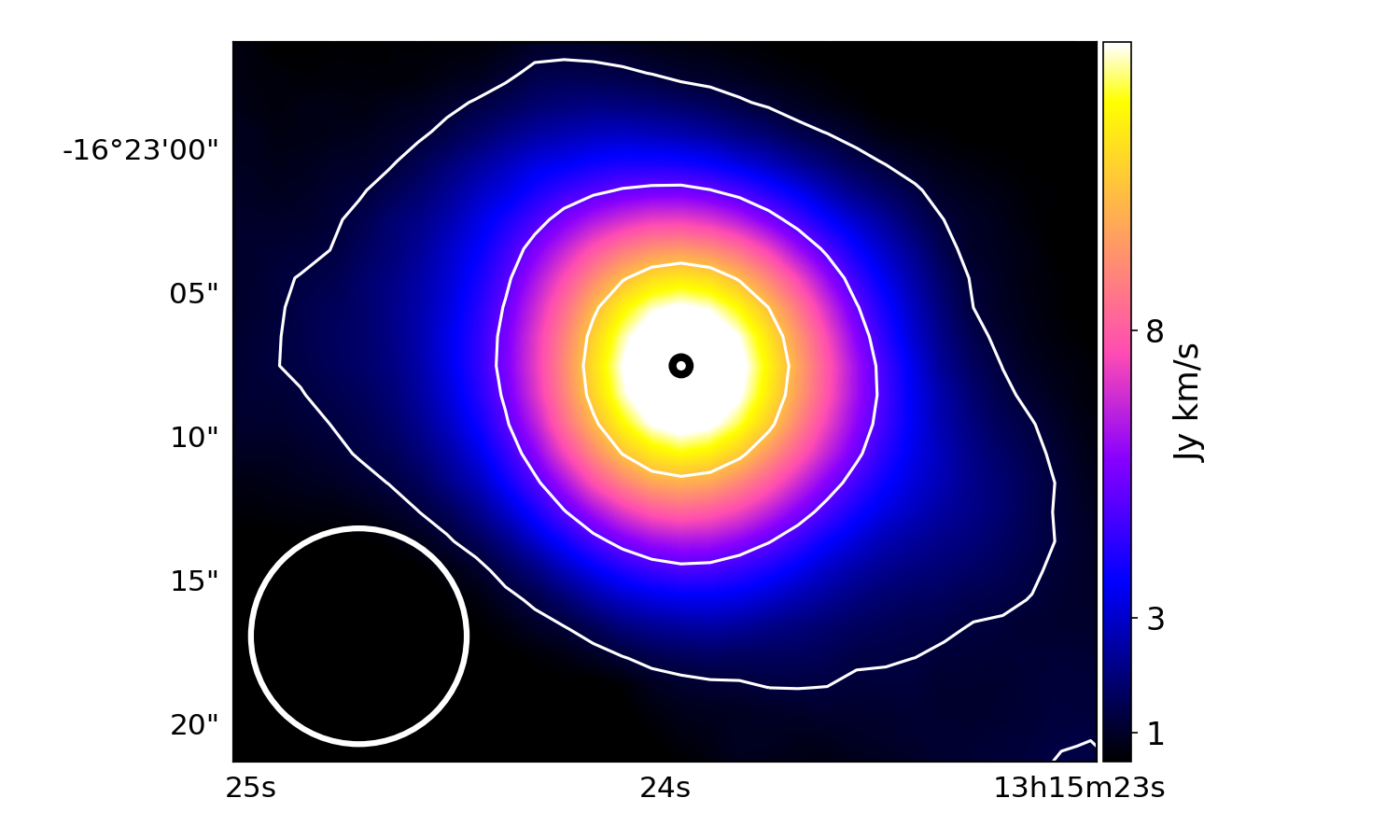}\\
\includegraphics[width=0.99\linewidth]{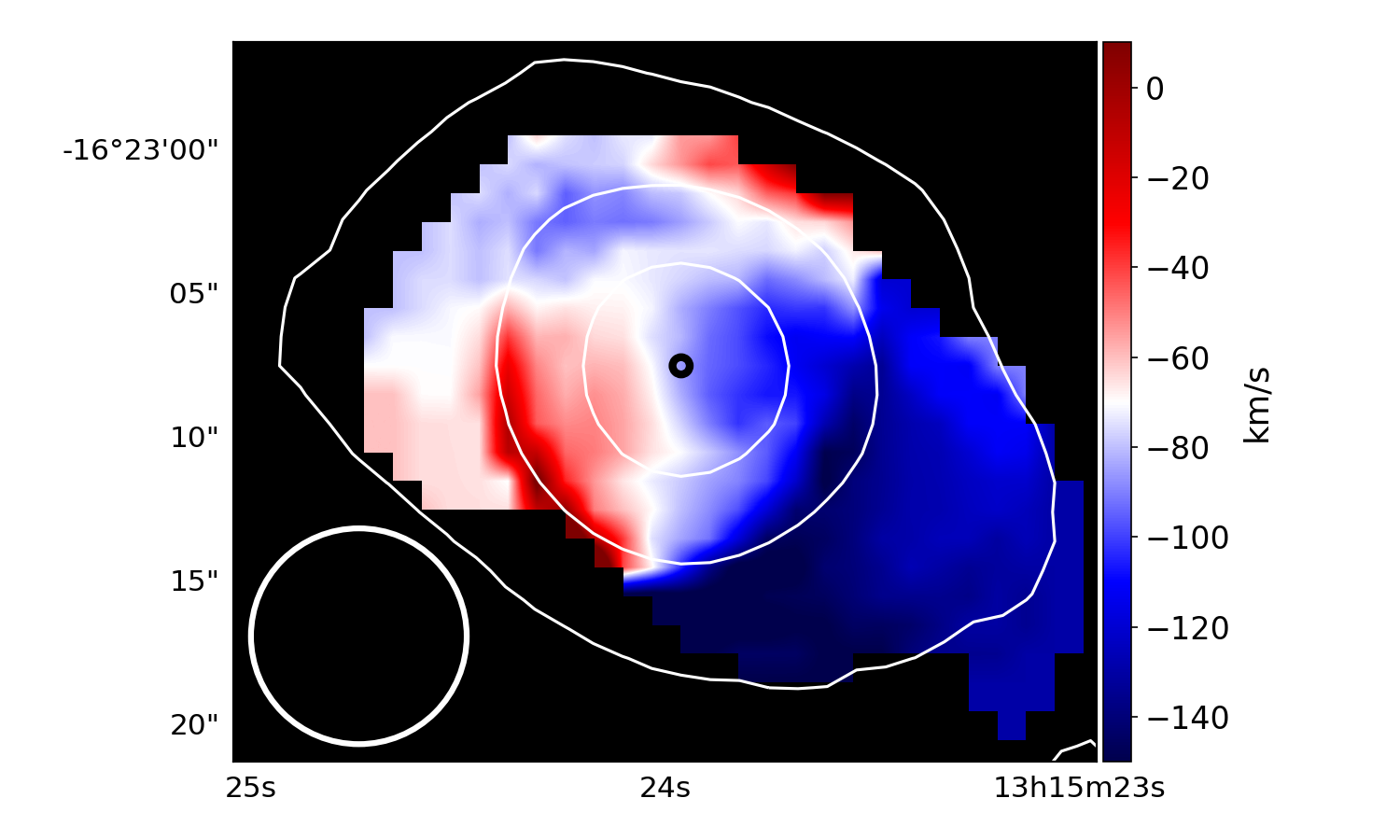}\\
\includegraphics[width=0.99\linewidth]{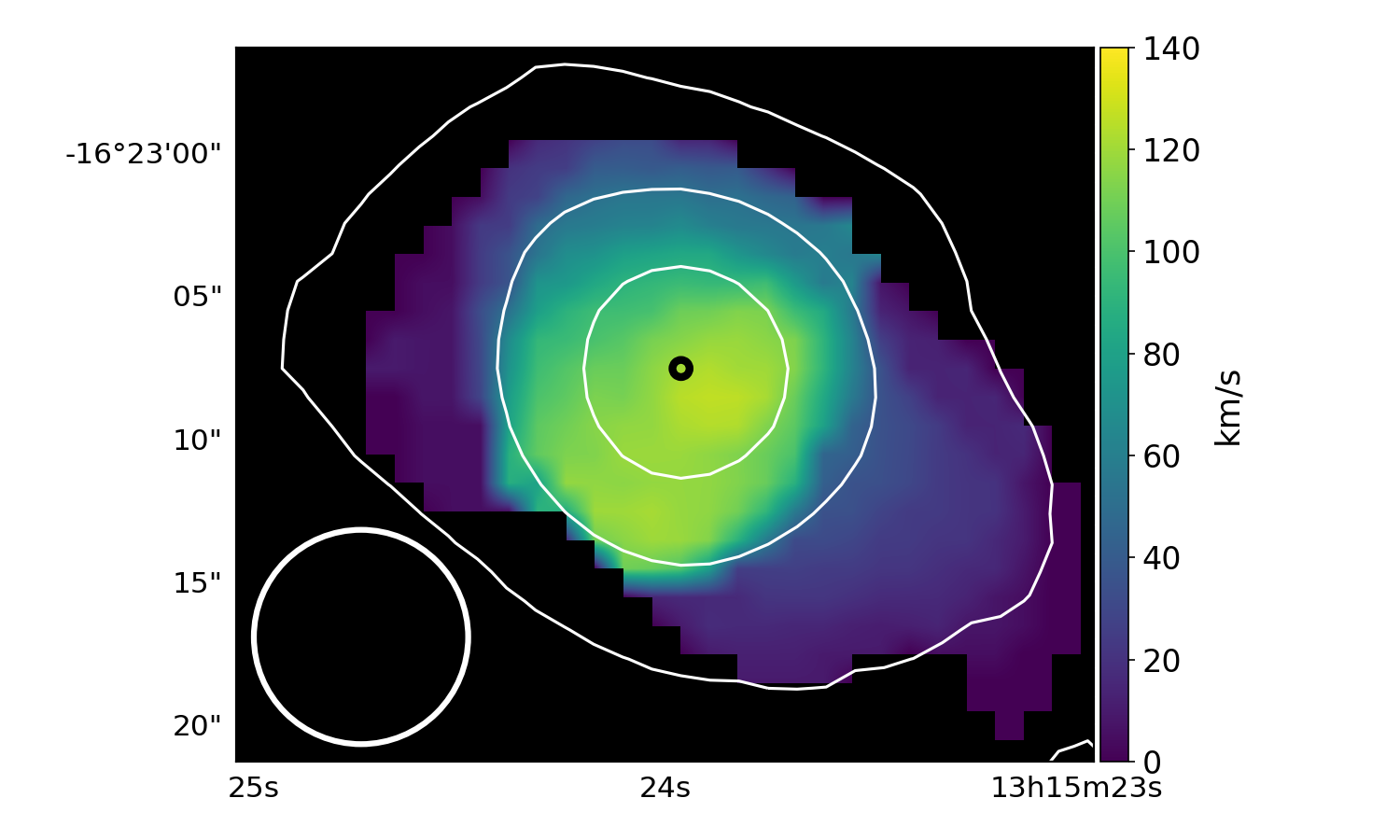}
\caption{Zeroth (flux, top), first (velocity, middle), and second (dispersion, bottom) moments of the ACA CO(2-1) emission between $\num{-550}$ and $\SI{300}{km\,s^{-1}}$.  
The first and second moments only include pixels with S/N greater than 5. White contours show the $3, 12, 24 \sigma$ levels of the zeroth moment map. The black circle in the center shows the location of the central AGN. The most redshifted emission is oriented along the diagonal north-east south-west axis.
\label{fig:moments}}
\end{figure}

\begin{figure*}
    \centering
    \includegraphics[trim=0px 150px 0px 0px,clip,width=0.9\linewidth]{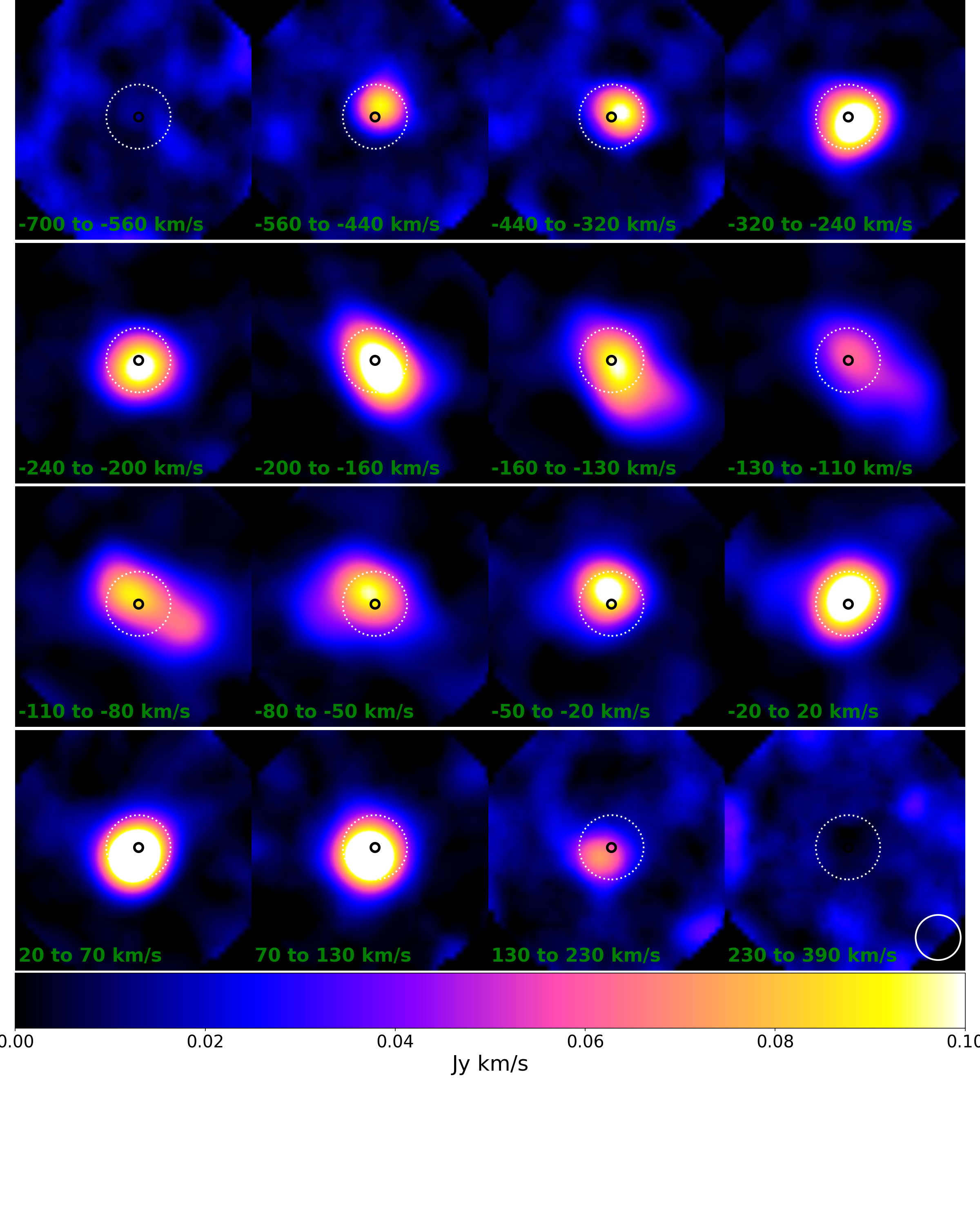}
    \caption{Zeroth moment maps of the ACA data cubes, binned to have the same integrated flux in each velocity bin. The black circle marks the position of the continuum source, while the white dotted circle marks the $6\arcsec$ ($\SI{900}{pc}$) radius of IRAM. Each image is about $\SI{40}{\arcsec}$ across. We find extended emission (clearly outside the IRAM beam) between $\num{-200}$ and $\SI{-50}{km\,s^{-1}}$. The ACA beamsize is shown in the lower right corner.}
    \label{fig:ACA_specbins}
\end{figure*}

We generated \nth{0} (integrated flux density), \nth{1} (velocity), and \nth{2} (velocity dispersion) moment maps (see Fig. \ref{fig:moments}) by integrating the CO(2-1) emission between $\num{-550}$ and $\SI{300}{km\,s^{-1}}$.
The \nth{1} and \nth{2} moment maps only include pixels with a $\mathrm{S/N}\geq5$, while the \nth{0} moment map includes all pixels above $\mathrm{S/N}=1$. 
We also mask out regions where the primary beam response of the 7m telescopes drops below 50\%.
Due to the moderate ACA beam, the only deviation from spherically symmetric emission in the \nth{0} moment map is a slight extension toward the southwest (see also velocities $\num{-160}$ to $\SI{-80}{km\,s^{-1}}$ in Fig. \ref{fig:ACA_specbins}). 
The peak of the CO(2-1) emission is coincident with the location of the AGN. 

A large scale velocity gradient is evident in the \nth{1} moment map (Fig. \ref{fig:moments}) along an east-west direction, with the most highly blueshifted emission located toward the southwest (i.e., the extended feature seen in the \nth{0} moment map in Fig. \ref{fig:moments}). 
Most of the redshifted emission resides in two structures, a structure toward the southeast (coincident with cloud 18 in \citealp{David2014-jn}) and a fainter structure toward the northwest. 
The $2^\mathrm{nd}$ moment map (Fig. \ref{fig:moments}) shows that the velocity dispersion of the CO(2-1) emission increases from $\SI{10}{km\,s^{-1}}$  at large radii, up to  $\SI{140}{km\,s^{-1}}$ near the AGN, but this is mostly a projection effect, since the emission near the AGN is summed over many individual molecular structures along the line-of-sight. 
At large radii, there are probably only a few individual molecular structures along the line-of-sight, and the \nth{2} moment map shows that the velocity dispersion of these structures is less than $\SI{20}{km\,s^{-1}}$.

\subsection{Total Molecular Mass}\label{ch:total_mol_mass}
The molecular mass was computed assuming a Galactic conversion factor from \cite{Bolatto2013-oe},
\begin{equation}
    X_\mathrm{CO} = \SI{2e20}{cm^{-2}\,(K\,km\,s^{-1})^{-1}}~,
\end{equation}
which gives
\begin{equation}
    M_\mathrm{mol} = \num{1.05e4} ~ \frac{S_\mathrm{CO} ~\Delta v}{\si{Jy\,km\,s^{-1}}} ~\left(\frac{D_L}{\si{Mpc}}\right)^2 (1+z)^{-1} ~\si{M_\odot}~,
    \label{eq:X}
\end{equation}
\noindent
where $S_\mathrm{CO} ~\Delta v$ is the integrated CO (1-0) flux density and $D_L$ is the luminosity distance.
As in \cite{David2014-jn} we assume a CO(2-1) to CO(1-0) flux density ratio of $\num{3.2}$.
We note that Eq. \ref{eq:X} includes the contribution of molecular hydrogen and helium, while all other molecules (including CO) are negligible in terms of mass.  
While $X_\mathrm{CO}$ is known to depend on a number of factors (e.g., metallicity and other environmental factors), the only measurement of $X_\mathrm{CO}$ appropriate for molecular gas that has condensed out of a cooling flow
was done by \cite{Vantyghem2017-ym}, by observing both $^{13}$CO and $^{12}$CO in RXJ0821+0752.  Their results suggest that the molecular mass computed assuming a Galactic $X_\mathrm{CO}$ may overestimate the true mass by a factor of two. \cite{Lim2017-ig} also pointed out that the physical conditions of the molecular gas in cluster or group central galaxies are likely very different from those of giant molecular clouds in our Galaxy, raising concerns about using the same $X_\mathrm{CO}$ for both. Nevertheless, to provide a simple comparison with past work as well as across different group or cluster central galaxies, we apply the factor in equation \ref{eq:X}.
\begin{figure*}
    \centering
    \includegraphics[width=0.8\linewidth]{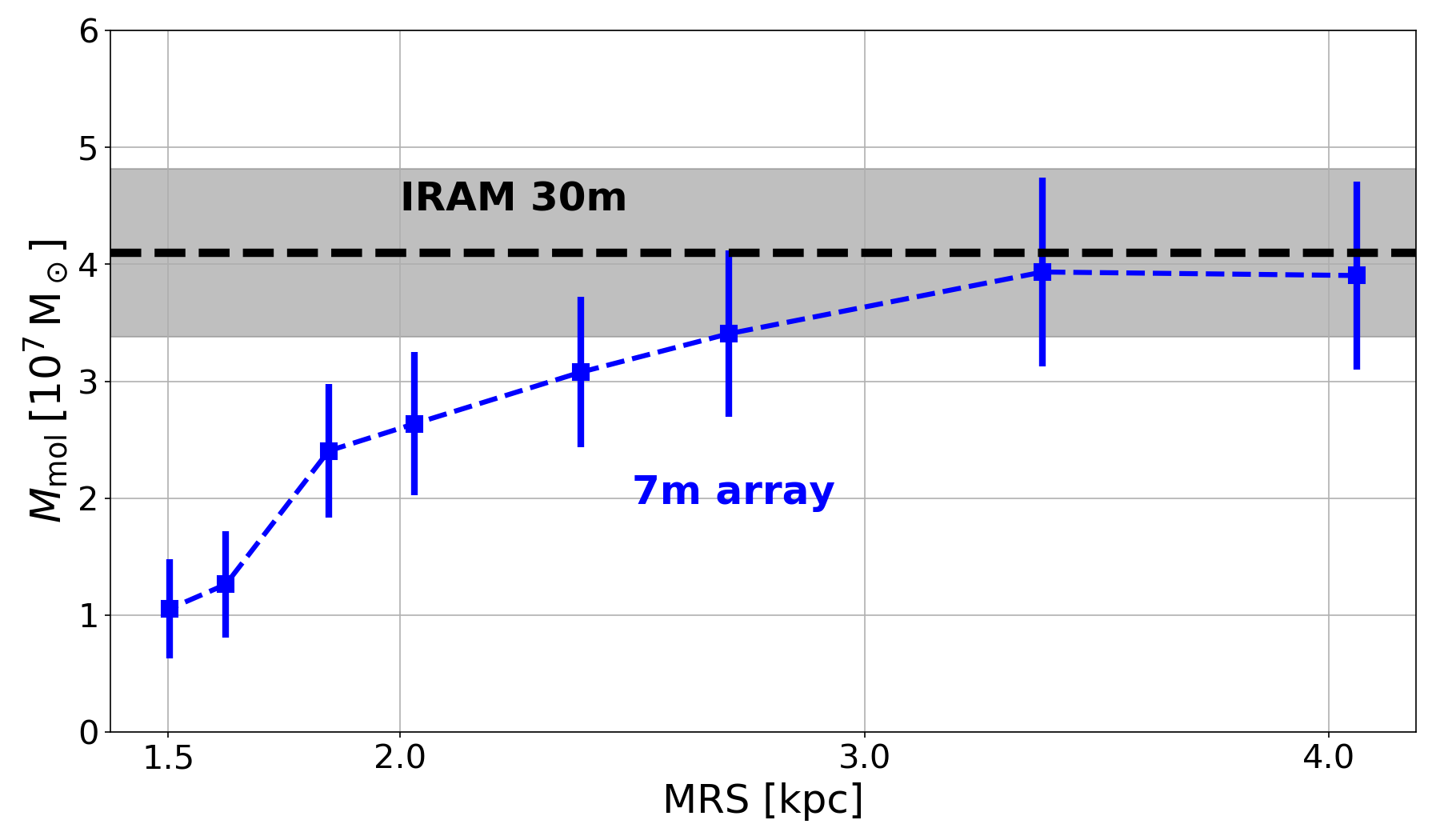}
    \caption{Total molecular mass as a function of the maximum resolvable scale (MRS) of the ACA 7m array CO(2-1) spectra extracted from images generated with different filtering on the smallest baseline.  All spectra are extracted from within $\SI{6}{\arcsec}$ radius apertures. The total IRAM signal is recovered from molecular clouds with scales below \SI{3.5}{kpc}. \label{fig:mass_mrs}}
\end{figure*}

\noindent
We computed the total molecular mass within a $\SI{15}{\arcsec}$ radius aperture (a physical radius of $\SI{2.3}{kpc}$) using three methods: 1) from the three Gaussian fits shown in Fig. \ref{fig:30asec_spec}, 2) from the \nth{0} moment image shown in Fig.\ref{fig:moments}, and 3) a masked-moment map (\citealp{Dame2011-hk}).
These methods yield total molecular masses of $\SI{10.3(2)e7}{M_\odot}$, $\SI{9.5(12)e7}{M_\odot}$, and $\SI{10.0(1)e7}{M_\odot}$, respectively, which are in relatively good agreement. 

In addition to determining the concentration of the molecular gas toward the center of NGC~5044, we also computed the total molecular mass within the $\SI{6}{\arcsec}$ radius aperture (a physical radius of $\SI{0.9}{kpc}$) shown in Fig. \ref{fig:12asec_spec} and obtained a (projected) mass of $\SI{4.2(1)e7}{M_\odot}$. 
For comparison, the deprojected (i.e., spherical) mass of the hot X-ray emitting gas within 0.75 and $\SI{1.5}{kpc}$ is $\SI{1.1e7}{M_\odot}$ and $\SI{5.6e7}{M_\odot}$, respectively. 
Thus, the molecular gas is more centrally concentrated than the  hot gas and dominates the total gas mass within the central kpc.

\begin{figure}
    \centering
    \includegraphics[width=0.99\linewidth]{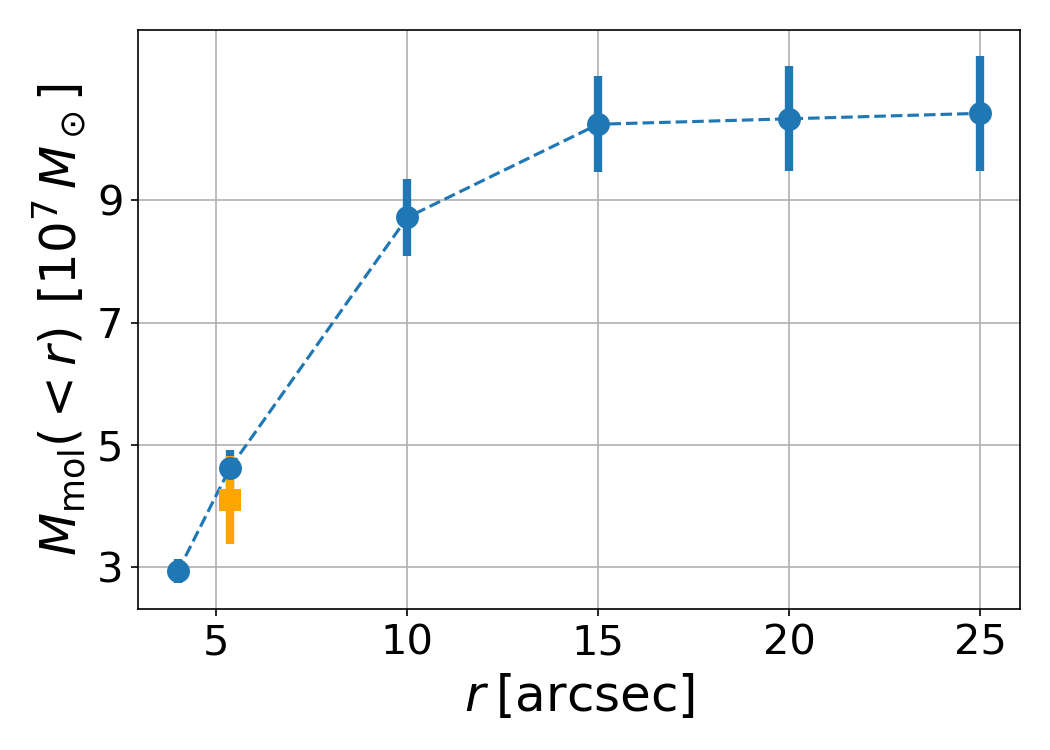}
    \caption{ACA cumulative molecular mass profiles (blue dots). The orange square represents the IRAM measurement. We see no CO(2-1) emission outside 1.5\,kpc radius (10\,arcsec).}
    \label{fig:TP_density}
\end{figure}

Interferometry enables us to test the spatial scales on which the CO emission is distributed: We filter the data on different uv scales to exclude the largest scales, and derive the mass only from smaller physical scales. The dependence shown in Fig. \ref{fig:mass_mrs} illustrates that most of the emission comes from scales of 2\,kpc, and the maximum size is $\SI{3}{kpc}$. Between 2 and 3\,kpc, the molecular mass is doubled.

\subsection{ACA Channel Maps}\label{ch:channelmaps}
A set of 16 channel maps from ACA (7m + TP) is shown in Fig. \ref{fig:ACA_specbins}.
Molecular structures are detected in all velocity bins between $\num{-560}$ and $\SI{230}{km\,s^{-1}}$. 
Due to the large ACA beam, most of the structures are only marginally resolved, but there are clearly two separate structures in the $\SI{-100}{km\,s^{-1}}$ velocity bin.  
The highly blueshifted structure is clearly detected in the $\num{-560}$ to $\SI{-440}{km\,s^{-1}}$ velocity bin, just $\SI{2.5}{\arcsec}$ to the north of the AGN. This velocity is about two times higher than the stellar velocity dispersion (see \citealp{David2009-hn}), and could be related to uplift of material by the AGN. 
An infalling cloud with $\sim \SI{500}{km\,s^{-1}}$ line of sight velocity has to be very close to the AGN to reach this velocity. However, the projected distance ($\SI{375}{pc}$) as a lower limit for the distance to the AGN excludes the possibility of an infalling cloud.

All of the blueshifted CO(2-1) emission between  $\SI{-200}{km\,s^{-1}}$ and 0 was resolved out in our higher resolution cycle 0 observation (\citealp{David2014-jn}).  
CO(2-1) emission can only arise from the cores of GMCs where the densities are comparable to the critical H$_2$ density of $\SI{e4}{cm^{-3}}$. All of the structures shown in Fig. \ref{fig:ACA_specbins} have a very low volume filling factor of molecular gas with most of the volume filled with hot, X-ray emitting gas.  

These results show that long baselines detect molecular structures with high GMC volume filling factors, or equivalently, regions where the GMCs are strongly clustered, and short baselines detect emission from molecular structures with low GMC volume filling factors, or where the GMCs are widely separated.  
Our results show that for nearby systems like NGC~5044, a broad range of baselines is required to determine the complete census of the molecular content.

The 7m array data combined with the TP observations allow us to measure the CO emission out to $\SI{20}{\arcsec}$ radius, which is close to the TP field of view. Figure \ref{fig:TP_density} shows the total molecular mass inferred from the intergrated CO(2-1) signal as a function of radius. We see no significant increase in emission beyond $\SI{15}{\arcsec}$ radius. 

\subsection{Molecular Cloud Properties}\label{ch:cloud_properties}
In order to identify individual clouds from the ACA signal, we used the \verb|CLUMPFIND| algorithm (\citealp{Williams1994-uc}). We use the ACA 7m array datacube with a $\SI{10}{km\,s^{-1}}$ velocity binning. 
We tested the detection algorithm with several velocity binnings from $\SI{2}{km\,s^{-1}}$ to $\SI{50}{km\,s^{-1}}$. 
The very fine binning of $\SI{2}{km\,s^{-1}}$ yields a large number of clouds, most of them just in a single or two velocity bins, and thus not useful for us to derive further properties. 
The difference between $\SI{5}{km\,s^{-1}}$ and $\SI{10}{km\,s^{-1}}$ is small (9 vs. 8 clouds). A binning of $\SI{20}{km\,s^{-1}}$ and larger results in only a few clouds, which are not well sampled by the velocity resolution.
The 3D locations of the three largest clouds in the $\SI{20}{km\,s^{-1}}$ image cube are similar to the detections in the $\SI{10}{km\,s^{-1}}$ cube.
The optimal binning should be a compromise between a slight oversampling of clouds in velocity space and still having good sensitivity in the individual binned channels. If we assume an ideal oversampling in velocity space of a factor of 2 to 5 with respect to the cloud FWHM, we find that all the detected clouds are best covered by the $\SI{10}{km\,s^{-1}}$ binning, which will be our default for the cloud detection.

To know how many false detections are expected in our sample of clouds, we inverted the image cube (by multiplying with $-1$), and reran the detection algorithm on the new image cube, which should contain noise and possible absorption lines in the positive values. From this test we find two clouds (at 255 and $\SI{265}{km\,s^{-1}}$), which turn out to be real absorption features, that we discuss in more detail in section \ref{ch:absorption}. We do not find any false detection. 

Following \cite{Pineda2009-ir}, we adopt the optimal threshold and stepsize parameters for  \verb|CLUMPFIND|, $5\sigma$ and $3\sigma$, respectively, which results in 8 detected clouds (see Table \ref{tab:clouds}, Figs. \ref{fig:cloud_moments}, \ref{fig:cloud_moments2} showing the moment maps all except the smallest cloud).
The $5\sigma$ threshold makes false detections from noise unlikely.
Note that we also list the unidentified emission, i.e., everything outside the clouds, as ID 0.
In contrast to the SNR of features in the spectrum (as in Tab. \ref{tab:fit30arcsec} and \ref{tab:absorption}) the SNR of clouds (Tab. \ref{tab:clouds}) is the peak flux value of a cloud divided by the rms, which reflects the definition of them\verb|CLUMPFIND| threshold parameter.
The velocity width of cloud ID 8 cannot be determined since almost all pixels belonging to this clump were found in one single channel.
The summed masses of the individual clouds (ID 1-8), which are almost entirely within $\SI{8}{\arcsec}$, is $\SI{5e7}{M_\odot}$. This is about 70\% of the total molecular mass within a radius of $\SI{8}{\arcsec}$.
The location and mass of cloud ID 1 are in relatively good agreement with cloud 18 and 19 from \cite{David2014-jn} using only the 12m cycle 0 ALMA data. This cloud would have to consist of clumps below $\SI{600}{pc}$ size in order to be fully detected in the cycle 0 data. Other individual clouds from \cite{David2014-jn} cannot be found in the 7m data, due to the spatial resolution of the ACA data and low fluxes of these clouds, i.e. the flux density sensitivity is too low in the ACA data.

We also create individual moment maps (\nth{0}, \nth{1}, and \nth{2} moment) for each cloud by only selecting the pixels of the image cube assigned to a cloud by \verb|CLUMPFIND| (shown in Figures \ref{fig:cloud_moments} and \ref{fig:cloud_moments2}). For this we extended slightly (by a few arcseconds) the identification in the spatial dimensions to avoid sharp cutoffs. 
Clouds ID 2 and 5 show signs of rotation in the first moment map, while ID 6 seems to have an irregular, non-circular shape.  The bottom panel of Fig. \ref{fig:cloud_moments2} shows moments of the unclassified emission (everything except detected clouds), which generally appears as a ring around the central regions where most clouds are detected. This is likely a selection effect, since clouds at larger projected distances are usually not as bright, and the instrumental sensitivity is reduced. 
We investigate the possibility of rotating clouds in section \ref{ch:clouds}.

\begin{deluxetable}{cccccc}
 \tablecaption{Cloud properties}
 \tablehead{
 \colhead{ID} &
 \colhead{$\left\langle v \right\rangle$} &
 \colhead{$\sigma$} &
 \colhead{M$_{mol}$} &
 \colhead{$r$} & 
  \colhead{SNR}
 \\
 \colhead{} &
 \colhead{(km~s$^{-1}$)} &
 \colhead{(km~s$^{-1}$)} &
 \colhead{($10^7$~M$_{\odot}$)} &
 \colhead{($\arcsec$)} & 
 \colhead{}
 }
\startdata
1 & \phm{-}\phn 60\phd $\pm$ 5   & \phn21 $\pm$ 10 & 1.16 $\pm$ 0.12 & 2.2 & 19\\
2 & \phn\phn-31\phd $\pm$ 15 & 21 $\pm$ 8 & 1.19 $\pm$ 0.11 & 2.0 & 15\\
3 & -172 $\pm$ 7  & 10 $\pm$ 6 & 0.74 $\pm$ 0.06 & 2.8 & 13\\
4 & \phn-109 $\pm$ 12 & 16 $\pm$ 9 & 0.81 $\pm$ 0.08 & 1.4 & 12\\
5 & -216 $\pm$ 6  & 10 $\pm$ 6 & 0.53 $\pm$ 0.05 & 2.0 & 12\\
6 & \phn-124 $\pm$ 10 & 10 $\pm$ 7 & 0.37 $\pm$ 0.04 & 7.6 & 10\\
7 & \phn-269 $\pm$ 12 & \phn 16 $\pm$ 15 & 0.24 $\pm$ 0.04 & 2.2 & 8\\
8 & -542 $\pm$ 4 & $<5$ & $<0.01$ & 2.2 & 5\\
0 & \phn\phn-174 $\pm$ 153 & 277 $\pm$ 96 & 3.34 $\pm$ 0.18 & 1.0 & \\
\enddata
\tablecomments{Velocity centroid, velocity dispersion, molecular mass, and projected distance from the continuum source of the detected clouds. ID0 corresponds to all emission not included in detected clouds. \label{tab:clouds}}
\end{deluxetable}

\subsection{Comparison with clouds in ALMA Cycle 0 data}
We detect 8 molecular gas clouds in our ACA dataset with the \verb|CLUMPFIND| algorithm. The Cycle 0 dataset of the ALMA 12m array ($\sim\,$6 times higher spatial resolution, see Table \ref{tab:obslognew}) has been analyzed by \cite{David2014-jn} who detected 24 clouds.
Although the two datasets are sensitive to clouds on different spatial scales, we can make a comparison between the detected clouds (see Table \ref{tab:comparison}).
\begin{deluxetable}{cccc}
 \tablecaption{Comparison between Cycle~0 and ACA detected clouds}
 \tablehead{
 \colhead{ACA} &
 \colhead{Cycle 0} &
 \colhead{C0 $\left\langle v \right\rangle$} &
 \colhead{$f_\mathrm{C0}$} \\
 \colhead{Cloud ID} &
 \colhead{ Cloud ID} &
 \colhead{(km~s$^{-1}$)} &
 \colhead{(\%)}  \\
 \colhead{(1)} &
 \colhead{(2)} &
 \colhead{(3)} &
 \colhead{(4)}
 }
\startdata
1 & 18 & 59 & 87 \\
2 & 13 & -38 & 28\\
3 & 6, 7, 8 & -160 & 80\\
4 & 9 & -113 & 10\\
5 & 2,3,4 & -230 & 89 \\
6 & 10 & -132 & 25
\enddata
\tablecomments{Columns are the detected cloud in the ACA data (1), see also Table \ref{tab:clouds}), the corresponding cloud detections in the Cycle 0 data (2), from \cite{David2014-jn}, the average velocity (3) and the fraction of the ACA cloud mass found in corresponding Cycle 0 clouds (4).      \label{tab:comparison}}
\end{deluxetable}

\cite{David2014-jn} found 11 clouds in the Cycle 0 data which are located at a large offset angle from the continuum source. We exclude these clouds from the following considerations since the noise level at large offset angles (2.5\,kpc) is about three times larger than in the phase center of the same observation. This can lead to false detections by the \verb|CLUMPFIND| algorithm.


Two clouds from the Cycle 0 data could not be clearly identified in the ACA image cube (Cycle 0 ID 11 and 15), although they are located very close to the center. 
Cycle 0 cloud ID 15 has the lowest mass of all the detected clouds ($\SI{3e5}{M_\odot}$), so it is likely below the detection threshold of the ACA data. Cloud 11 has a large molecular mass ($\SI{4e6}{M_\odot}$), but also a very broad velocity distribution causing the flux in the $\SI{10}{km\,s^{-1}}$-channels to be below the ACA sensitivity (which is worse than the Cycle 0 threshold). 

We also calculate for each ACA detected cloud the fraction of molecular mass that was detected as a Cycle 0 cloud (Column 4 in Table \ref{tab:comparison}). As seen already in the spectrum (Fig. \ref{fig:12asec_spec}), the fluxes of ACA clouds 1, 3 and 5 have been almost entirely captured by the Cycle 0 observation, while the other clouds are more extended and cannot be fully detected by the ALMA 12m array.

\subsection{Absorption features in the AGN spectrum}\label{ch:absorption}
\begin{deluxetable}{cccccc}
 \tablewidth{\linewidth} 
 \tablecaption{Gaussian fits to the AGN absorption features.\label{tab:absorption}}
 \tablehead{
 \colhead{Velocity} &
 \colhead{FWHM} &
 \colhead{Amplitude} &
 \colhead{Mass} &
 \colhead{Radius} & 
 \colhead{SNR}
 \\
 \colhead{(km~s$^{-1}$)} &
 \colhead{(km~s$^{-1}$)} &
 \colhead{(mJy)} &
 \colhead{($\SI{e3}{M_\odot}$)} &
 \colhead{(pc)} & 
 \colhead{}
 }
\startdata
$\num{255.5(3)}$ & $\num{6.6(3)}$ & $\num{9.5(8)}$ & 7 & 4 & 15\\
$\num{264.7(4)}$ & $\num{6.3(4)}$ & $\num{7.2(8)}$  & 6 & 4 & 9\\
\enddata
\tablecomments{The two Gaussian fits to the 12m spectrum extracted within an aperture representing the beam size (center in Fig. \ref{fig:absorption}). Masses and radii have been estimated from the linewidth, assuming the velocity-radius relation, and that the system is in virial equilibrium.}
\end{deluxetable}

\begin{figure*}
\centering 
\includegraphics[width=0.99\linewidth]{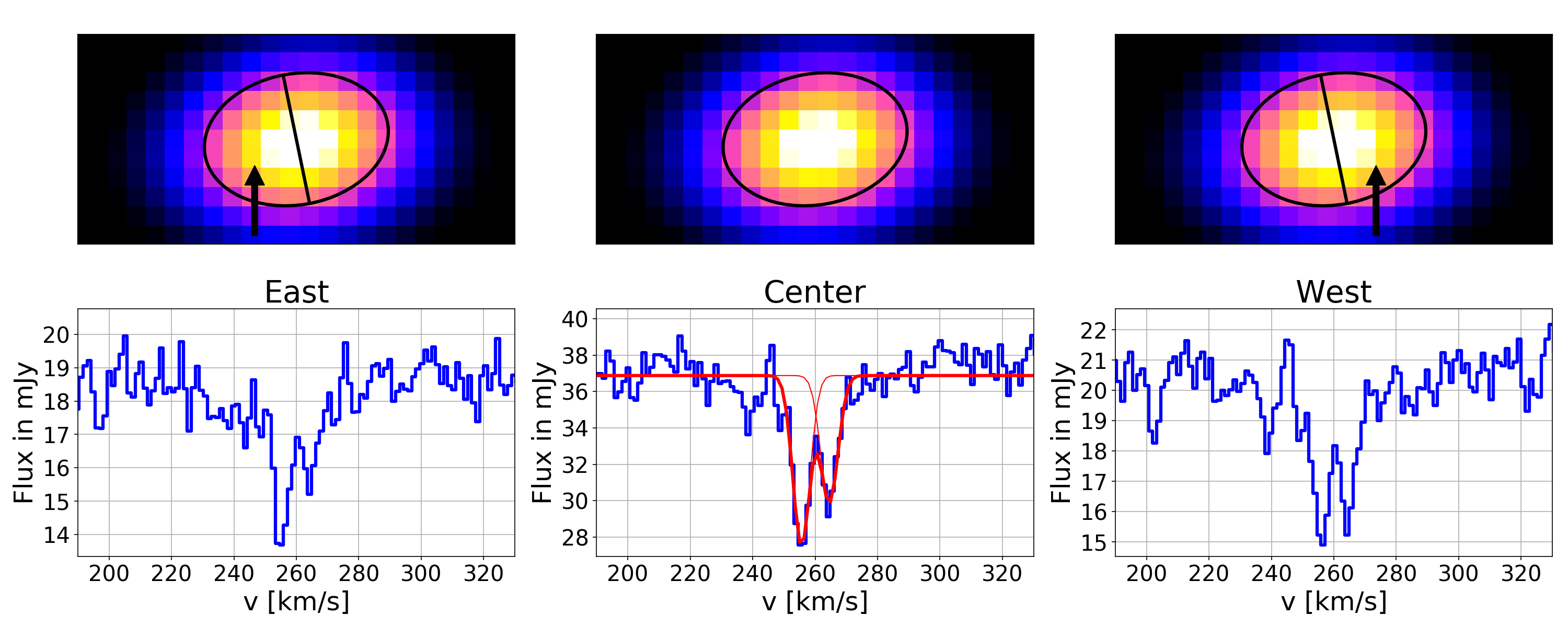}
\caption{Spectrum of the central continuum source highlighting the two strong, redshifted absorption features at $v=255$ and $\SI{265}{km\,s^{-1}}$. The feature at higher redshift is strongest in the western part of the continuum source. In the center panel the aperture is centered on the continuum source and has the extent of the beam, while in the other two cases the eastern or western half is excluded. The spatial dependence of the absorption feature indicates that the continuum source is only slightly smaller than the PSF.
}
\label{fig:absorption}
\end{figure*}

In section \ref{ch:cloud_properties} we noted the detection of two absorption features with respect to the emission of the AGN in the inverted (negative) image cube. 
We analyse the ALMA cycle 4 extended array data (see Tab. \ref{tab:obslognew}) with superior spatial resolution to characterize the nature of this absorption feature. 

We detect two absorption features at a velocity of $\sim\SI{260}{km\,s^{-1}}$ in the continuum subtracted CO(2-1) spectrum (Fig. \ref{fig:absorption} shows the absorption in the continuum spectrum for three apertures) at the location of the central continuum source (AGN). 
The proximity of the absorption features to the AGN, at least as seen in projection, can be estimated from the extent of the $\SI{230}{GHz}$ continuum emission. 
We note that \cite{David2014-jn} detected one absorption line in the continuum spectrum of the ALMA cycle 0 data of NGC~5044 with a velocity of $\SI{260}{km\,s^{-1}}$, and a linewidth of $\SI{5.2}{km\,s^{-1}}$. 

To obtain the best spatial resolution, we imaged the 12m cycle 4 extended array observation using a Briggs robust parameter of -2, selecting only baselines longer than $\SI{1000}{k\lambda}$, and a velocity bin of $\SI{2000}{km\,s^{-1}}$ (to maximize the S/N). 
This produced a restoring beam of $\SI{0.08}{\arcsec}$ by $\SI{0.05}{\arcsec}$, which corresponds to a geometric mean radius of $\SI{5}{pc}$. 
Using the CASA task \verb|imfit|, we found that the resulting \nth{0} moment image of the AGN continuum emission at $\SI{230}{GHz}$ is consistent with that of a point source. 

The absorption features can be modeled by Gaussians, for which best-fit values are given in Table \ref{tab:absorption}. As for the emission spectrum (Tab. \ref{tab:fit30arcsec}) the SNR values reported here are integrated within the FWHM.
Following \cite{Rose2019-tu} we can derive the molecular hydrogen column density of the absorbing clouds from the optical depth profile, $\tau_{ul}$, 
\begin{equation}
    N_\mathrm{tot} = Q(T_\mathrm{ex}) \frac{8 \pi \nu_{ul}^3}{c^3} \frac{g_l}{g_u} \frac{1}{A_{ul}} \frac{1}{1 - e^{-h\nu_\mathrm{ul} / k T_\mathrm{ex}}} \int \tau_{ul} \mathrm{d} \nu~ .
\end{equation}
$Q = 15.52$ is the partition function for a given excitation temperature, $T_\mathrm{ex}$ (we assume $\SI{42}{K}$, as derived by \citealp{Rose2019-tu} for Hydra A), $g_l=3$ and $g_u=5$ are the  degeneracies of the lower and upper level, $A_{ul} = \num{6.914e-7}$ is the Einstein coefficient, and $\nu_{ul}=\SI{230.538}{GHz}$ is the frequency of the CO(2-1) transition. For the absorption features in NGC~5044 we obtain a CO column density of $\SI{2.2e17}{cm^{-2}}$, which converts to a H$_2$ column density of $N_\mathrm{H_2} = \SI{1.5e20}{cm^{-2}}$ (\citealp{Sofia2004-ck}). This assumes a surface filling factor of 1, and the column density will be higher if the cloud does not fill the whole source.
Looking at all the ALMA observations between 2012 and 2018 we do not detect any significant change in the absorbing column density.

We notice that the amplitude of the second absorption line (centered at $\SI{265}{km\,s^{-1}}$) varies slightly with position, mainly along the east-west direction. 
This is illustrated in Fig.~\ref{fig:absorption} by the absorption spectra in three apertures along the east-west line. 
The amplitude of the second absorption line in the eastern half is about 3 times above the spectral noise level, while in all other cases the absorption lines are at least 5 times above the noise. 

With the release of \textit{CASA 5.5.0}\footnote{\url{https://casa.nrao.edu/casadocs/casa-5.5.0/introduction/release-notes-550}} the \verb|perchanweightdensity| parameter was introduced for the \textit{tclean} task. It allows the calculation of the Briggs weighting scheme on a per-channel-base for image cubes. Changes are predominantly small, however the second absorption line in the eastern half of the continuum source changes: Setting this parameter to \verb|True| increases the amplitude of this line, erasing any significant differences between the two absorption lines in the western or eastern half of the continuum source. Without deeper, high resolution data, unfortunately, we are unable to confirm the existence of a spatially variable absorption feature. 


\section{Discussion}\label{ch:discussion}
\subsection{Gas cooling}
\cite{Werner2014-vw} detected [CII] emission to a radius of $\SI{8}{kpc}$, which is associated with atomic/nuclear material (since NGC~5044 lacks significant star formation, see \citealp{Werner2014-vw}). 
We detect molecular CO(2-1) emission out to about $\SI{2}{kpc}$ (the TP observations are sensitive to more than $\SI{3}{kpc}$), while it is possible that a more extended component of CO is hidden below our detection threshold. Assuming there is a connection between the atomic and molecular gas components, a more extended molecular component at a level comparable to the rms in the \nth{0} moment maps and with a uniform extent over $8\times 8\,\si{kpc^2}$, can increase total molecular mass to $\sim\SI{e8}{M_\odot}$.



\subsubsection{Feedback models}
In contrast to Bondi accretion, the chaotic cold accretion (CCA) model (e.g., \citealp{Gaspari2013-eq,Voit2015-op,Tremblay2016-xg,Gaspari2017-ev}) predicts the condensation of cold clouds out of the turbulent atmosphere. These clouds will eventually ``rain'' on the SMBH, and trigger feedback. 

The simulations for the CCA model (\citealp{Gaspari2018-ni}) include hydrodynamics, cooling, heating, and, most importantly, turbulence at sub-pc resolution. Predictions include small clouds with large line-of-sight velocities and small velocity dispersions (thus larger volume filling factors $\sim 2\%$) close to the SMBH. Molecular gas with smaller line-of-sight velocities and larger dispersions are located at larger distances to the central AGN. In any case, turbulent pressure is key to prevent clouds from collapsing. Any significant turbulence will also prevent high star formation rates.

Prior studies using deep, high-resolution X-ray data have shown that the hot gas within the central region of NGC~5044 has been perturbed by at least three cycles of AGN outbursts and the motion of the central galaxy within the group potential. 
Cavities at various distances from the AGN have been related to the oldest outburst (\citealp{Giacintucci2011-ru}), intermediate outbursts (\citealp{David2009-hn}), and the most recent AGN activity cycle (\citealp{David2017-ig}).
The mechanical heating by AGN-inflated cavities is sufficient to offset radiative cooling of the gas within the central $\SI{10}{kpc}$. 

A contrasting feedback description by \cite{McNamara2016-ux} claims that feedback from the AGN is stimulated through buoyantly rising cavities which lift low entropy gas, with molecular clouds forming in the wakes of the uplifted cavities (``updraft model''). Significant turbulence should exist in the wakes of the rising bubbles leading to density fluctuations and cooling as discussed by \cite{Gaspari2013-eq}. 
\cite{Brighenti2015-cu} suggest that those cavities trigger episodes of enhanced cooling.
As shown in Fig. \ref{fig:Halpha} the cavities (red) are aligned in NW-SE direction, while the H$\alpha$ filaments and the CO are elongated perpendicular to this axis. The updraft model makes a clear link between the cold, molecular gas and the rising bubbles, and one would expect a spatial correlation between the two. 

\subsubsection{H$\alpha$ in NGC~5044}
\begin{figure*}
\begin{center}
\includegraphics[width=0.85\linewidth]{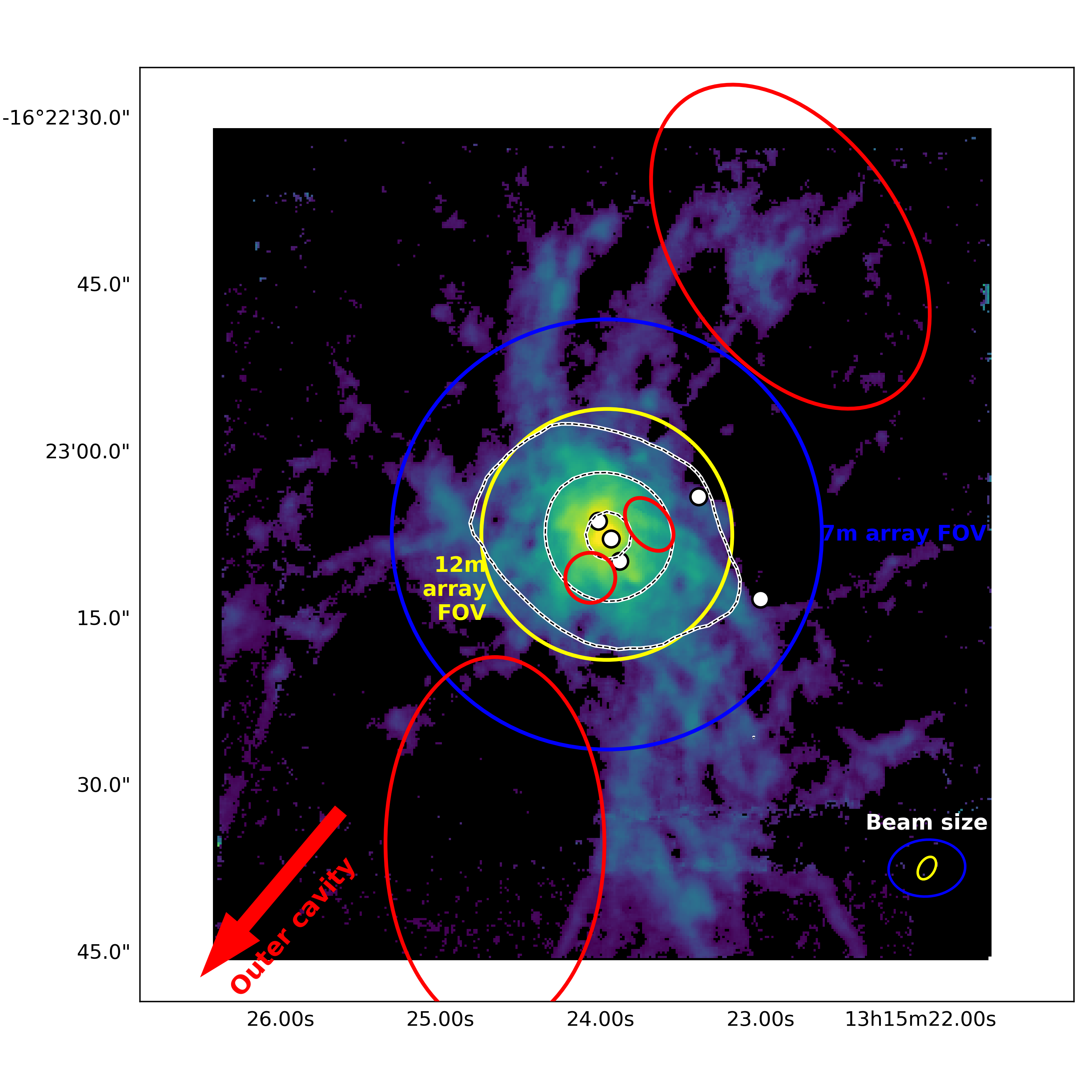}
\caption{MUSE H$\alpha$ image (Sun et al., in prep) of NGC~5044 with CO contours. White/black-dashed contours show the detected CO emission from ACA. Red ellipses indicate the detected cavities in the X-ray image (\citealp{David2017-ig}), and white-filled circles mark the locations of the resolved CO clouds in the ALMA cycle 0 data (\citealp{Temi2018-rr}). The major axis orientation of the CO emission is roughly consistent with the H$\alpha$ filament. Blue and yellow circles/ellipses show the 7m and 12m array FOVs and beam sizes. \label{fig:Halpha}}
\end{center}
\end{figure*}
To draw a link between the cavities and cold gas we have to take into account the distribution of warm, line-emitting ionized gas in NGC~5044 traced by $H\alpha$ emission. 
NGC~5044 was observed with MUSE in four nights from Jan. 17, 2015 to Feb. 3, 2015, for seven exposures of 820 sec each (PI: Hamer). MUSE provides a spectroscopic data cube on a rectangular $\SI{1}{\arcmin}\times\SI{1}{\arcmin}$ field, with a spectral coverage of $\SIrange{4800}{9000}{\angstrom}$. 
The MUSE data were reduced using the v2.6.2 ESO MUSE pipeline  within the Zurich Atmosphere Purge (ZAP) software (\citealp{Soto2016-ux}). More details on the MUSE data reduction will be given in a subsequent paper (Sun et al., in prep.).
 

\begin{figure*}
\begin{center}
\includegraphics[height=200pt]{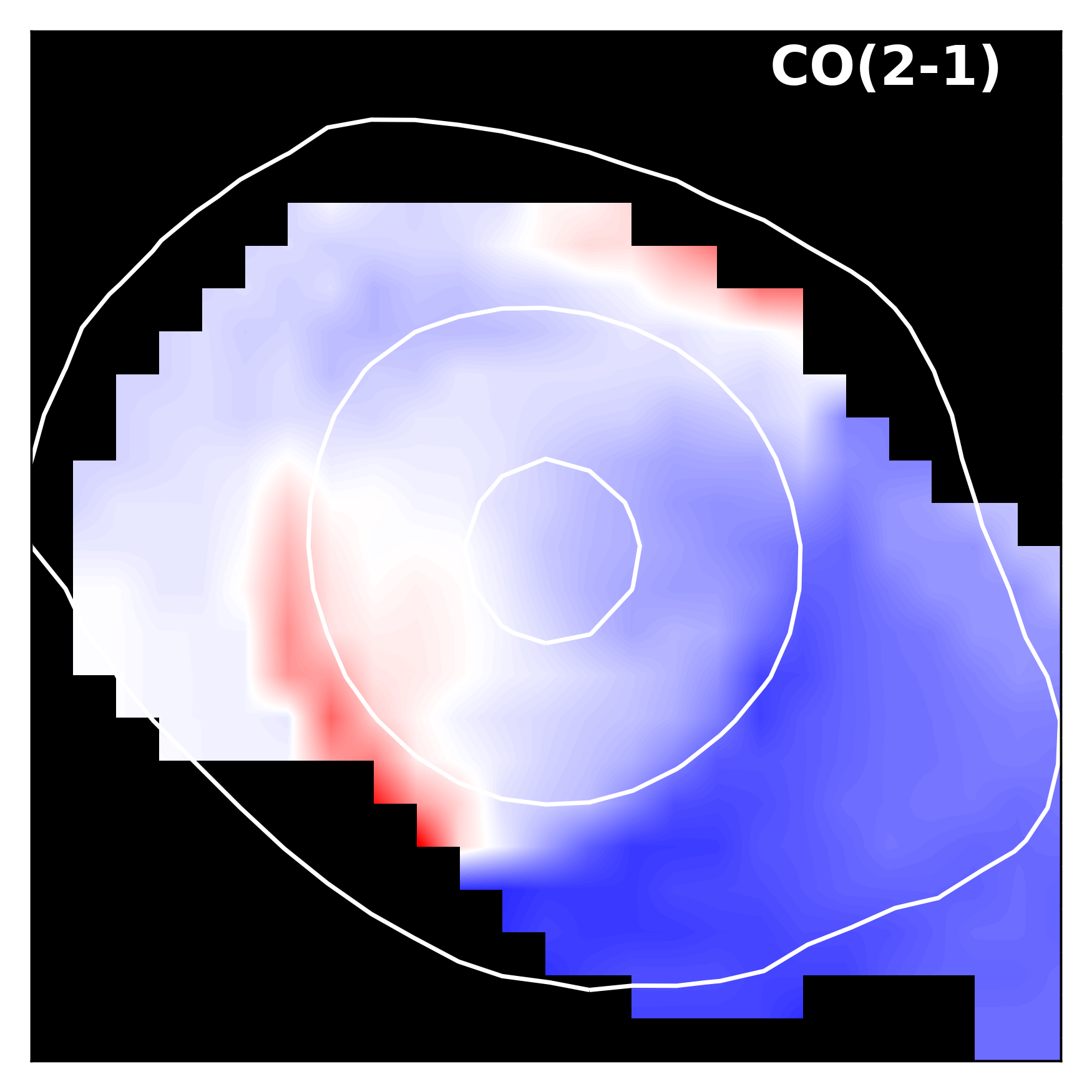}
\includegraphics[height=200pt]{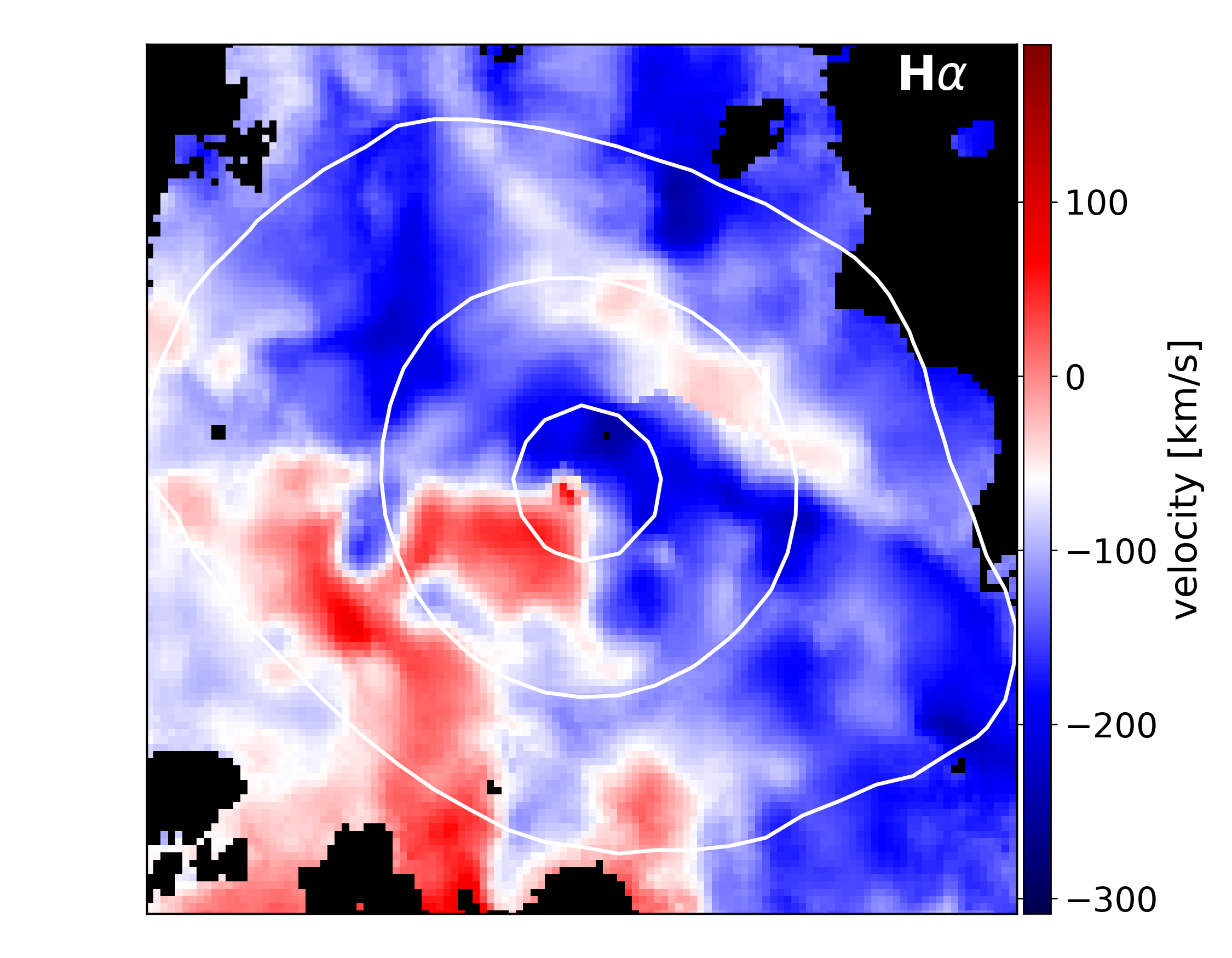}
\caption{CO (2-1) velocity map close up (left, as in Fig. \ref{fig:moments}), and MUSE H$\alpha$ \nth{1} moment map (Sun et al., in prep) of NGC~5044 with CO contours (right). The image size is $\SI{24}{\arcsec}$ ($\SI{3.6}{kpc}$). \label{fig:Halpha_vel} }
\end{center}
\end{figure*}
Here, we include the MUSE H$\alpha$ image (Fig. \ref{fig:Halpha}) and the central velocity map (Fig. \ref{fig:Halpha_vel})  to compare with the ALMA data.

We find an S-shape distribution in H$\alpha$ observations. Within the central 3\,kpc the cold CO emitting gas is aligned well with the H$\alpha$ filaments along a NE-SW axis (white/black-dashed contours in Fig. \ref{fig:Halpha}). Resolved CO clouds (\citealp{Temi2018-rr}) in the center are aligned along the same axis, while some clouds are located at the outer edge of the H$\alpha$ filaments (filled white circles in Fig. \ref{fig:Halpha}). 
At larger distances, we do not detect CO emission and the H$\alpha$ filaments are bent toward the N-S direction.
The cavities are distributed along a NW-SE axis (red ellipses in Fig. \ref{fig:Halpha}), which seems to be uncorrelated with the CO emission. 
We note that an intermediate-distance cavity in the south is located between the H$\alpha$ filaments, and seems to be partly surrounded by it. This could be an indication that there is a connection between the cavities and the H$\alpha$ filaments.
However, we also see molecular clouds in other region of the central part of the galaxy, unrelated to cavities. 
Based on our finding we cannot exclude either, the CCA or the updraft model for the feedback processes in NGC~5044.

The velocity structure of the H$\alpha$ filaments shown in Fig. \ref{fig:Halpha_vel} shows a trend of blueshifted gas to the NW and redshifted emission to the SE. 
In the central 2\,kpc (Fig. \ref{fig:Halpha_vel}), we find some similarities between the CO and the H$\alpha$ filamental velocity structure, although the resolution of the two maps is very different.

\subsubsection{Comparing the molecular clouds in NGC~5044 with CCA model predictions}
\label{ch:cca}
\begin{figure}
    \centering
    \includegraphics[width=0.5\textwidth]{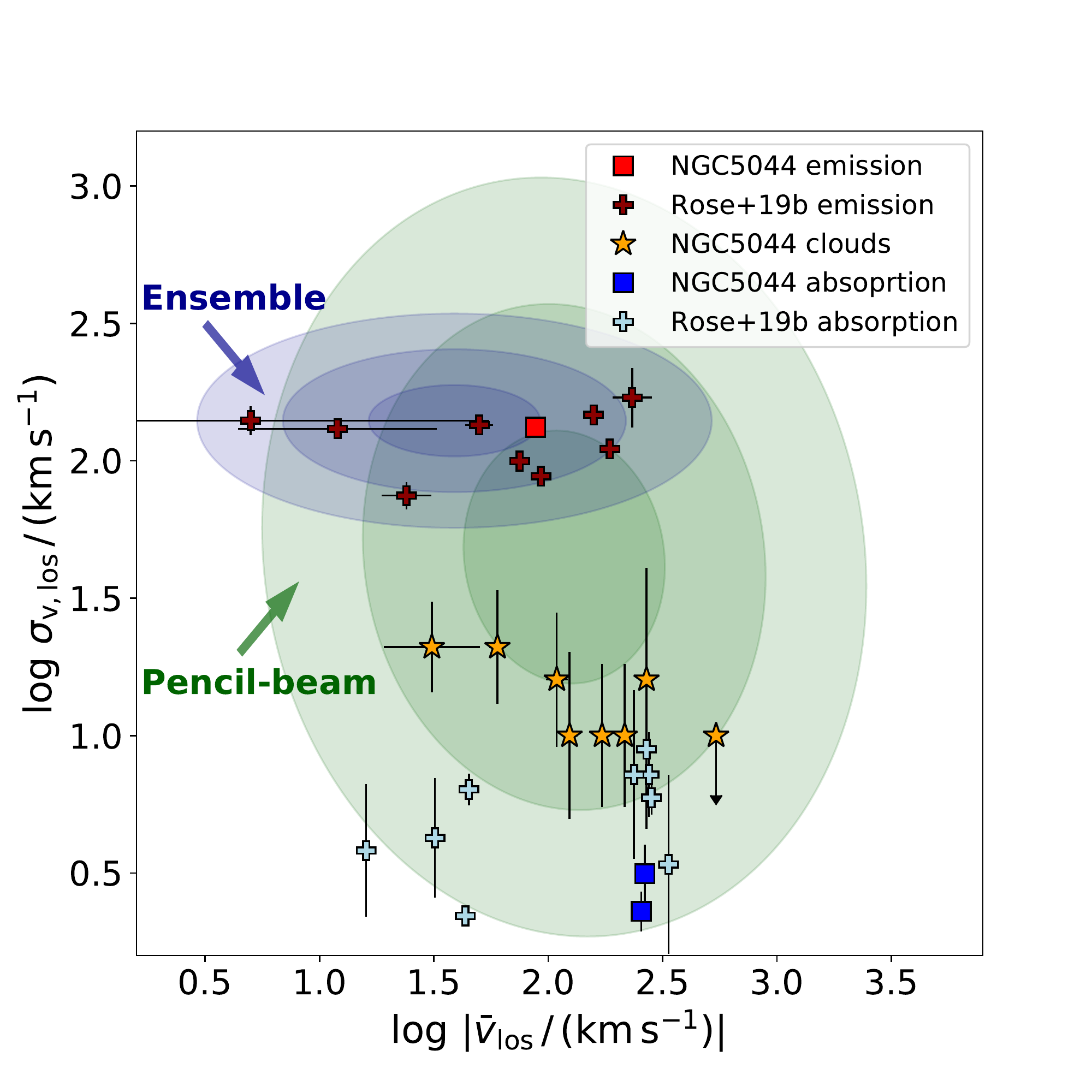}
    \caption{Comparison of observations with the CCA model (\citealp{Gaspari2018-ni}). The blue shaded region shows single spectrum/ensemble predictions, while the green shaded ellipse is associated with pencil-beam detections in simulations. Datapoints show observational detections: CO emission in red, CO absorption in blue, and detected clouds in orange. \label{fig:gaspari}}
\end{figure}

Simulations of the CCA model predict molecular gas to be in a certain location of the velocity shift -- velocity dispersion diagram (Fig. \ref{fig:gaspari}). Those two quantities trace the kinematics of molecular structures depending on the detection aperture. Either one traces an ensemble in a larger aperture excluding the nuclear region (blue ellipse in the top left of Fig. \ref{fig:gaspari}), or a small ``pencil-beam'' aperture trough the center of the AGN (green ellipse in the center of Fig. \ref{fig:gaspari}).
The ensemble region comprises larger amount of condensed gas at larger distances from the AGN, while the pencil beam typically picks up single small (low velocity dispersion) clouds with large line-of-sight velocity components close to the AGN.

All emission and absorption datapoints (blue and red crosses and squares in Fig. \ref{fig:gaspari}) are detected features in the CO spectra. 
Only orange stars in Fig. \ref{fig:gaspari} show identified clouds.
We find that the ACA detected clouds (orange stars, Tab. \ref{tab:clouds}) are within the expected range of the pencil-beam detections, while the expected range for ensemble-beam detections (single spectrum approach) is shifted to higher velocity dispersion and smaller velocity offset. 

We also overplot the CO emission found by \cite{Rose2019-au} for a few nearby objects ($z < 0.2$) detected from the CO spectrum (dark red crosses), which all fall in the ensemble region. 
The absorption detections from \cite{Rose2019-au} (light blue crosses) and our absorption features (blue squares) populate the lowest part of the plot, just within the pencil-beam detection region. The bias of these small clouds seen in absorption toward lower $\sigma_\mathrm{v, los}$ could be caused by a lower level of turbulence in the hot gas.

Our ACA clouds (orange stars) are the only detections found through a cloud detection algorithm. This explains their location in Fig. \ref{fig:gaspari} between the large molecular filaments detected through spectral fitting, and the small clouds seen in absorption. Our detection for NGC~5044 from the ACA spectrum (in this case using only a single Gaussian to be consistent with the ensemble assumption) shown as a red square in Fig. \ref{fig:gaspari}) are shifted to higher velocity dispersions, consistent with the CCA ensemble prediction (blue ellipse). 
In sum, we find consistent results with the predictions of the CCA model. However, our \verb|CLUMPFIND| detected clouds in the ACA data fill the gap (in the velocity dispersion -- line-of-sight velocity phase space) between larger ensembles and small individual clouds found in absorption.

\subsection{Total amount of molecular gas}
The hot gas temperature within the central $\SI{15}{\arcsec}$ (2.3\,kpc) in NGC~5044 is about $\SI{0.8}{keV}$. 
We adapt the power-law model of the hot gas density distribution from Eq. (2) in \cite{David2017-ig} to derive the hot gas mass within 1.5\,kpc of $\SI{3e7}{M_\odot}$, which is about 50\% of the molecular mass within the same radius.
The molecular gas in NGC~5044 is about an order of magnitude lower than that found in the Perseus cluster (\citealp{Lim2008-fa,Salome2008-kr}). 
Other galaxy groups have been found to host more molecular gas than NGC~5044 (\citealp{OSullivan2018-yk}), likely due to merger, but NGC~5044 contains more molecular gas than any other cooling flow group we know of (i.e., a combination of a short central cooling time and an X-ray bright cool core). 
A possible explanation for this is that the AGN in NGC~5044 is currently not in an active phase, and the galaxy group has not experienced recent mergers, allowing  the formation of molecular gas through the cooling processes in the center. 
The gas depletion timescale for the molecular gas in NGC~5044 is at least $\SI{800}{Myr}$, given the current upper limit for the star formation rate  estimated from a combination of FUV and MIR data ($\SI{0.073}{M_\odot\,yr^{-1}}$, \citealp{Werner2014-vw}), which is in agreement with central dominant galaxies in galaxy groups (\citealp{OSullivan2018-yk}). 
However, the duty cycle of the outbursts in NGC~5044 is about $\SI{e8}{yr}$, so most of the molecular mass will be reheated.

\subsection{Cloud Properties}\label{ch:clouds}

\subsubsection{Virial equilibrium in detected clouds}
\label{ch:vel_disp}
Knowing the sizes of the newly detected molecular clouds is important to derive other quantities.
We applied the CASA task \verb,imfit, to the \nth{0} moment maps of the individual clouds to measure the extent (beam deconvolved) from our ACA data. 
We can derive (beam deconvolved) sizes for clouds 1 to 5, where we define the size as the square root of the product of the two fitted FWHM along the major and minor axis. We find 
$\SI{308(22)}{pc}$,
$\SI{684(10)}{pc}$,
$\SI{717(16)}{pc}$,
$\SI{538(24)}{pc}$, and
$\SI{541(21)}{pc}$ for the first 5 clouds, all with axis ratios around 2, except cloud 2 which is almost circular. The other clouds are not resolved. 
Together with the cloud mass, $M$, and the velocity dispersion, $\sigma_\mathrm{v}$ (both given in Tab. \ref{tab:clouds}), we can test whether virial equilibrium is established using the expression for the virial parameter $\alpha$ (\citealp{Bertoldi1992-zf}),
\begin{equation}
    \alpha =  \frac{5 R \sigma^2_\mathrm{v}}{G M}~,
    \label{eq:sigma_size}
\end{equation} 
that gives the ratio of kinetic and gravitational energy. Perfect virial equilibrium is established for $\alpha=1$.
For the five resolved clouds, we find $\alpha$ of typically around 8. Due to the large uncertainties of the cloud velocity dispersions, these values are also very uncertain. However, we can derive an average, $\alpha = \num{8.6(34)}$. The ACA detected clouds are much closer to bound structures than smaller clouds detected in cycle 0 data  by \cite{Temi2018-rr} which have $\alpha$ ranging from 30 to 650 (250 on average). 
Due to its shorter baselines, the 7m array is able to detect large scale emission (see Fig. \ref{fig:arrays}), which traces the size of the clouds more accurately.
However, even values of $\alpha = 8$ are at the upper end of what has been found by simulations of molecular clouds in the Milky Way (\citealp{Dobbs2011-ds}), and those clouds are still considered globally unbound. 
The dispersion timescale, $\frac{r}{\sigma_\mathrm{v}}$, is on the order of a few Myr (up to $\SI{35}{Myr}$), which sets an upper limit for the lifetime of the clouds.

For two of the clouds (IDs 2 and 5), we find rotational characteristics in the \nth{1} moment maps, from which we estimate maximum velocity difference within each cloud by eye to be on the order of 20 to $\SI{40}{km\,s^{-1}}$.
This can be used to derive a dynamical mass, if  we assume rotational support. However, the typical timescale for one rotation is several times above the dispersion timescale. Therefore, we do not assume that the clouds are actually globally in rotation, but consist of smaller sub-clouds in motion.
The combination of the  7m array data with the higher resolution 12m array will enable us to resolve the clouds better and measure the average density and the radius. We will present this in a subsequent paper.

\subsection{Molecular clouds in absorption}
The spectrum of the central AGN, which is visible in the continuum image as a point source, shows two clouds in absorption (Fig. \ref{fig:absorption}).
The two clouds must reside within a projected distance of $\SI{5}{pc}$ to the AGN (resolution of the continuum image). 
For comparison, the sphere of influence of the central black hole, 
\begin{equation}
R_a=G M_\mathrm{bh}/\sigma_*^2 \approx \SI{17}{pc}~,
\label{eq:influence}
\end{equation}
where $M_\mathrm{bh}= \SI{2.3e8}{M_\odot}$ and $\sigma_* = \SI{237}{km\,s^{-1}}$, is more than three times larger than the upper limit of the projected distance of the two clouds from the black hole.
If the infalling clouds are actually located within the sphere of influence of the black hole, they will eventually likely be accreted.

\subsubsection{Absorbing cloud properties}
As described in \cite{Tremblay2016-xg} we can calculate the mass of an absorbing molecular cloud from the line width assuming virial equilibrium,
\begin{equation}
    M_\mathrm{cloud} \approx \frac{R_\mathrm{cloud} \sigma^2}{G}~.
    \label{eq:abs_cloud_mass}
\end{equation}
For the two absorption line features at $\num{255}$ and $\SI{265}{km\,s^{-1}}$, we determine masses around $\SIrange{6e3}{7e3}{M_\odot}$ (see Tab. \ref{tab:absorption}).
The expected size can be derived from the scaling relation by \cite{Solomon1987-yg}, 
\begin{equation}
    \sigma = (\num{1.0(1)}) \times S^{\num{0.50(5)}} \si{km\,s^{-1}}~,
    \label{eq:solomon}
\end{equation}
where the size $S$ is given in parsec. With Eq. \ref{eq:solomon} we find a cloud size of roughly a factor of 2 smaller than the observed (unresolved) upper limit on the central continuum source extent ($\SI{5}{pc}$ radius).
We can estimate, from the integrated absorption signal, the H$_2$ column density of the absorbing material to be roughly $\SI{2.3e21}{cm^{-2}}$. Given the estimated mass and size of the cloud, we obtain a density of $\SI{400}{cm^{-3}}$, which is comparable to that of a GMC. 
The line widths and densities of the clouds along the line of sight result in a turbulent pressure of about $\frac{P}{k_\mathrm{B}} = \SI{1.9e6}{K\,cm^{-3}}$, which is in agreement with the hot gas pressure of $\sim \SI{2e6}{K\,cm^{-3}}$ (\citealp{David2014-jn}). However, cloud mass and size estimates are based on scaling relations, so it is hard to conclude if the absorbing clouds are really in pressure equilibrium with the surrounding gas.
Following the argument by \cite{Field2011-kn} the external pressure of the hot gas implies a maximum mass of the the molecular cloud under the assumption of a stable equilibrium. For the absorbing clouds in NGC~5044 this maximum mass is about 30 times higher than the derived masses from Eq. \ref{eq:abs_cloud_mass}.

\subsubsection{Distance to the AGN}
\label{ch:distance}
\cite{Tremblay2016-xg} found CO absorption in A\,2597, and inferred the absorbing material location was close to the central black hole of A\,2597 (about 1\,kpc in projection). The authors also find, at a similar velocity, HI absorption from a VLBA observation (resolution about 20\,pc).
In NGC~5044 we find CO absorption at a projected distance of less than $\SI{5}{pc}$. 
If the real separation of the absorbing clouds from the AGN is close to the projected distance, we have detected molecular clouds much closer to the AGN than previously observed in a cooling flow system.

In the following we try to derive the real separation of the absorbing clouds from the AGN.
\cite{Bruggen2016-sh} show that various timescales have to be considered when estimating the lifetime of a molecular cloud in the environment of X-ray bright gas. The time after which a cloud is destroyed by a shock induced through the motion within the ICM is the cloud crushing timescale,
\begin{equation}
    t_\mathrm{cc}  = \left(\frac{T_\mathrm{ext}}{T_\mathrm{cloud}}\right)^{0.5} \frac{R_c}{v_h} \approx \SI{7}{Myr}~,
    \label{eq:tcc}
\end{equation}
where $T_\mathrm{cloud}\approx \SI{10}{K}$ (\citealp{Braine1992-or}) and $T_\mathrm{ext} \approx \SI{0.8}{keV}\,k_\mathrm{B}^{-1}$ (\citealp{David2017-ig}) are the cloud and surrounding gas temperatures, respectively, $R_c \approx \SI{2}{pc}$ is the cloud radius, and $v_h = \SI{260}{km\,s^{-1}}$ is the velocity of the cloud.
Thermal conduction will heat up the cold cloud within the evaporation time, 
\begin{equation}
    t_\mathrm{evap} = \frac{R_c}{v_h} \frac{100}{f(M)} \approx \SI{0.8}{Myr}~,
    \label{eq:tevap}
\end{equation}
where $f(M) = 1$. This timescale is a lower limit and can be longer in the presence of magnetic fields. Cooling of the cloud through radiation would also decrease this timescale, but is not considered here.
The third timescale to be considered is the Kelvin-Helmholtz instability timescale,
\begin{equation}
    t_\mathrm{KH} \approx  5 \frac{R_c}{v_h} \left[ 1 + 4(\gamma-1) \frac{v_h}{c_\mathrm{sc}} \right]^{0.5} \approx \SI{240}{Myr}~,
    \label{eq:tKH}
\end{equation}
where $\gamma = \frac{7}{5}$ is the adiabatic index for CO, and $c_\mathrm{sc} \approx \SI{6.6e-2}{km\,s^{-1}}$ is the sound speed in the cloud.
We conclude that the infalling cloud should last for at least $\SI{400}{kyr}$, if not destroyed by cloud-cloud collisions.


\begin{figure}
\centering 
\includegraphics[width=0.99\linewidth]{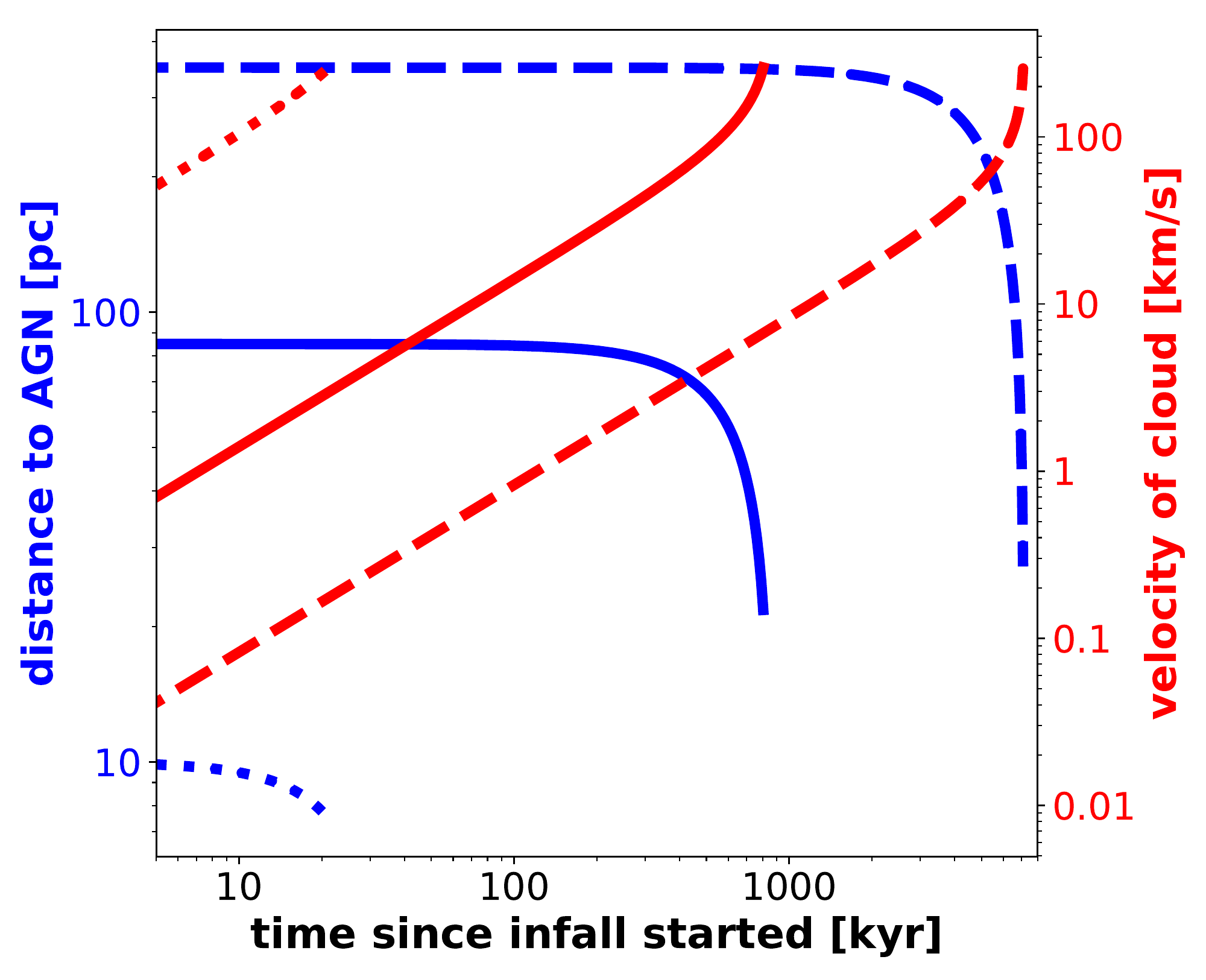}
\caption{Infall simulation showing velocity (red) and distance of the cloud to the AGN (blue) as a function of time. The different linestyles refer to initial distances of the cloud (350\,pc: dashed, 85\,pc: solid, 10\,pc: dotted). For each initial distance the cloud reaches the observed velocity ($\SI{260}{km\,s^{-1}}$) after different time (350\,pc: 7\,Myr, 85\,pc: 800\,kyr, 10\,pc: 21\,kyr), and at a different (final) distance to the AGN (350\,pc: 27\,pc, 85\,pc: 22\,pc, 10\,pc: 8\,pc).}
\label{fig:infall}
\end{figure}
Given the lifetime of the cloud we can calculate how close the cloud is to the AGN when reaching a relative velocity of $\SI{260}{km\,s^{-1}}$, assuming only gravity accelerates the cloud towards the supermassive black hole (SMBH).
As shown in Fig. \ref{fig:infall} a cloud, falling from a distance of $\SI{350}{pc}$ ($\SI{85}{pc}$, $\SI{10}{pc}$), will reach our threshold velocity at a distance of $\SI{27}{pc}$ ($\SI{22}{pc}$, $\SI{8}{pc}$) after $\SI{7}{Myr}$ ($\SI{800}{kyr}$, $\SI{21}{kyr}$). If we take the evaporation timescale as our threshold for the cloud lifetime, we conclude that the cloud is $\SI{22}{pc}$ or closer to the SMBH.
This is very close to the sphere of influence of the SMBH ($\SI{17}{pc}$, see Eq \ref{eq:influence}).
Assuming a uniform distribution of clouds, there is a $\frac{\SI{17}{pc}}{\SI{22}{pc}} \approx 80\%$ chance that the two clouds are within the sphere of influence of the black hole. Given all our assumptions (clouds with mass $\SI{e4}{M_\odot}$ initially at still, falling within the thermal conduction time), this means that the absorbing clouds in NGC~5044 are closer to the black hole than the clouds in A\,2597, but are also orders of magnitude smaller.

We note that in the scenario described above we assume a cloud falling onto the SMBH through gravity only. However, in the CCA model (see Section \ref{ch:cca} clouds share the same velocity of the hot gas until they dynamically separate from the hot gas after some time, and leaving less time for the cloud to fall toward the SMBH.

\subsubsection{Spatial variability of the AGN absorption features}
The NGC~5044 continuum source (AGN) is variable on timescales of years at mm wavelengths (Alastair Edge, private communication), implying an upper limit on the size of at least one component of the source of $\sim\SI{1}{pc}$. With our four ALMA/ACA observations (between 2012 and 2018) we estimate a linear decline of the continuum source flux at $\SI{229}{GHz}$ of $\SI{-1.6}{mJy\,yr^{-1}}$. 
Our ALMA CO(2-1) observations also allow us to place limits on the spatial extent of the AGN: At $\SI{244}{GHz}$ (highest frequency spectral window), the beam size is $70\times\SI{40}{mas}~(10.5\times\SI{6}{pc})$. Even with this small beam, the continuum source appears unresolved. 
The \textit{CASA} task \verb|imfit| is able to deconvolve the source and places constraints on the physical extent ($\num{10} \times \num{7}\,\si{mas}$, or $\num{1.5} \times \num{1.0}\,\si{pc}$, with 10\% and 90\% relative $1\sigma$ uncertainties along the major and minor axis, respectively).
\begin{figure*}
\centering 
\includegraphics[width=0.49\linewidth]{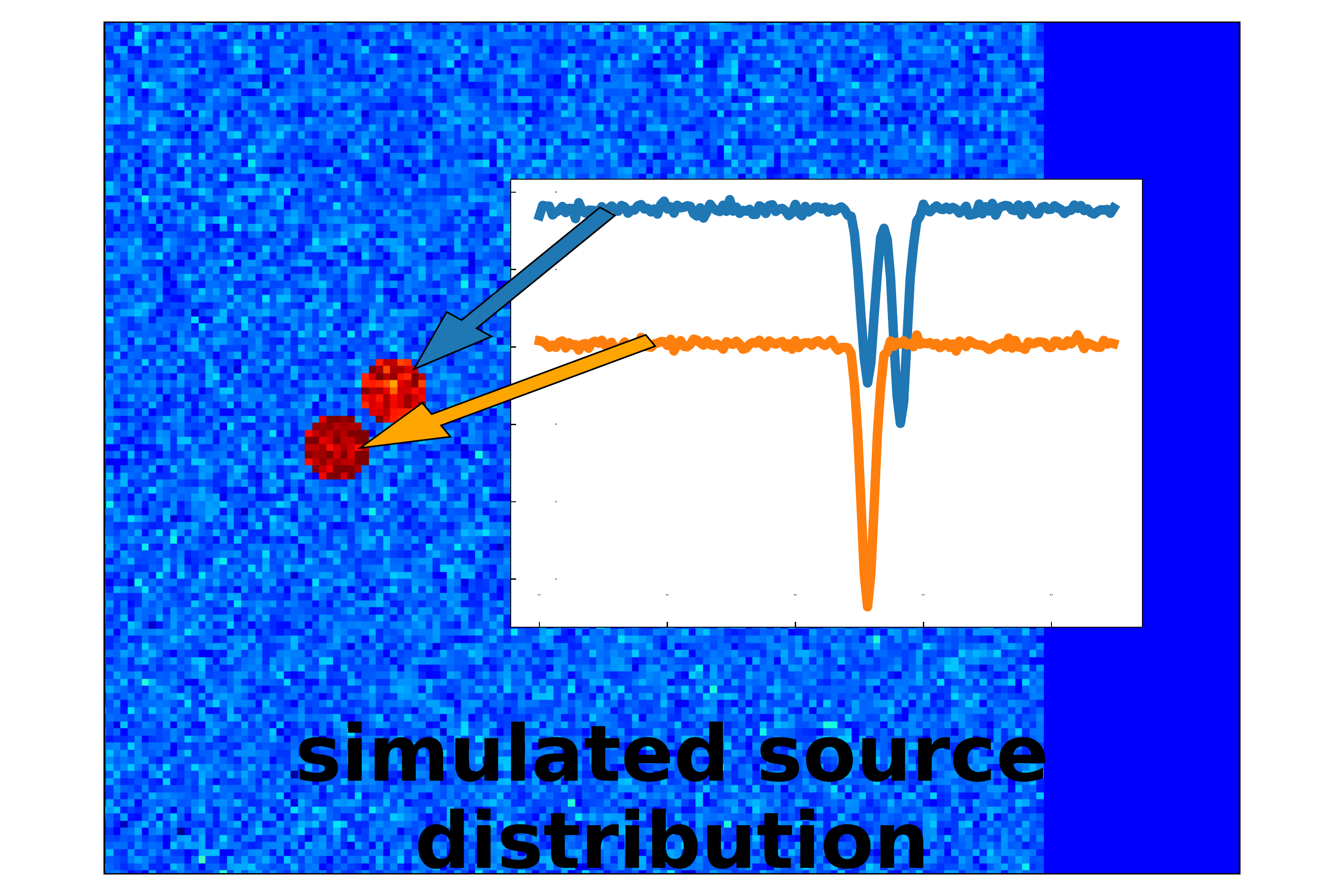}
\includegraphics[width=0.49\linewidth]{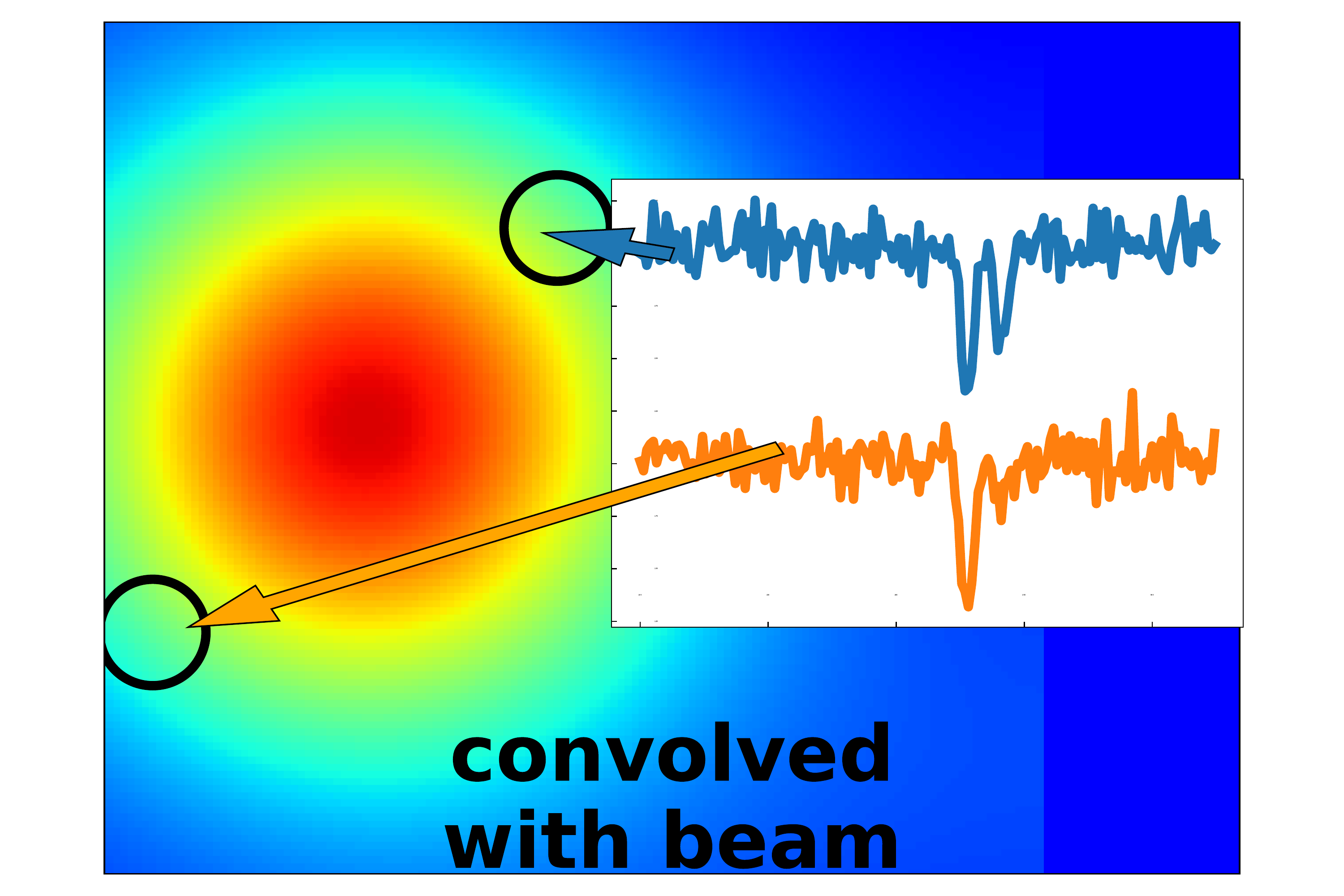}
\caption{\textit{Left:} Hypothetical distribution of two sources (red circles) with different spectra (shown in orange and blue). \textit{Right: } After convolution with a larger Gaussian only one point source is visible, but the spatial variability of the absorption spectrum (tested at two different points as shown by the black circles) is similar to our observations of NGC~5044.}
\label{fig:abs_sim}
\end{figure*}
The absorption features in the spectrum of the continuum source suggest spatial variability (see Section \ref{ch:absorption}), which cannot be confirmed with an alternative analysis mode. Deeper observations are needed to confidently confirm the weakened second absorption line in the western half of the continuum source. 

If the variability feature is real, we suggest two scenarios, (1) the CO emission of a cloud on the eastern side of the absorbing cloud partially covers the continuum source and fills (compensates) the absorption, and (2) the continuum source is extended near the ALMA resolution, and two different clouds cover two parts of the continuum source (e.g., core and jet). 
For (1), additional CO(2-1) emission is needed to fill in the absorption line on the eastern side of the core, where the 265\,km/s absorption line is less visible. The flux of this absorbed line would correspond to the emission of a cloud of $\sim \SI{1.4e5}{M_\odot}$. 
However, we can also estimate the mass from the line width of the absorbing cloud, assuming virial equilibrium, which results in a mass 50 times lower than the ``required'' mass, if emission compensates the absorption feature on one side. We also do not see any other signs of CO(2-1) emission in the continuum spectrum of the source, and scenario (1) is likely ruled out.

For the scenario (2), we estimated the feasibility with a basic simulation of two spatially separated continuum sources, each having an absorption feature at a different velocity (see Fig. \ref{fig:abs_sim} left panel). These two sources could represent a core and a jet. However, active jet spectra are typically steeper then the core spectra, which can make them hard to be detected at mm wavelengths. 
By smoothing the image cube spatially with a Gaussian about 3 times the separation of the two emitting continuum sources, we get an image consistent with a point source at the ALMA resolution (Fig. \ref{fig:abs_sim} right panel), which shows a very similar spatial dependence of the absorption as in NGC~5044. 

From the variability of the continuum source flux (on timescales of years), the unresolved nature of the point source at the highest frequency being not perfectly consistent with the beam shape, and the possibility of spatial variability of the absorption features, we have indications that the continuum source in NGC~5044 has a physical extent of a factor of a few below the ALMA resolution ($\SI{15}{pc}$). 

\subsubsection{Feedback energy}
NGC~5044 has the largest known reservoir of cold molecular gas of any dominant galaxy in a cooling-flow group, and three epochs of AGN activity are visible as cavities in its hot gas (\citealp{David2009-hn,Giacintucci2011-ru,David2017-ig}). In section \ref{ch:distance} we concluded that the two molecular clouds visible as absorption features in the spectrum of the continuum source are most likely within the sphere of influence of the AGN. We can now estimate the energy input to the AGN by those infalling clouds.

Within $\SI{800}{kyr}$ two clouds of total mass $\SI{1.3e4}{M_\odot}$ might fall onto the SMBH, which gives a mass accretion rate for this line of sight of $\SI{0.016}{M_\odot\,yr^{-1}}$. 
The possible feedback energy from the infalling cloud, assuming the mass - energy equivalence, is about $\SI{1.2e56}{erg}$, assuming a 50\% accretion efficiency and that 1\% of the available energy is turned into radiation or powering the jet.
This will be a lower limit for the accretion rate, since a shorter infall is possible. If the cloud is located at the sphere of influence, it covers a solid angle of about $\SI{80}{deg^2}$ as seen from the location of the black hole. Integrating over the whole sphere, the total mass accretion rate is $\SI{6}{M_\odot\,yr^{-1}}$, which is far above any typical mass accretion rate for central dominant galaxies, and it is about 550 times the Bondi accretion rate. This is an upper limit in case clouds are distributed everywhere around the SMBH.

For a distance close to the AGN ($\sim \SI{5}{pc}$) and a line of sight velocity ($\SI{260}{km\,s^{-1}}$), the free-fall timescale is about $\SI{140}{kyr}$, which implies a mass deposition rate of $\sim \SI{0.09}{M_\odot\,yr^{-1}}$. The energy rate (assuming again a radiative efficiency of $0.5\%$) is $\SI{2.6e43}{erg\,s^{-1}}$, which is several orders of magnitude higher than the bolometric X-ray or integrated radio luminosity of the AGN, and also 2 times higher than the total cavity power (southern cavity plus smaller northern cavities, see \citealp{David2009-hn}) which is related to several AGN outburst in the past. This implies that more than half of each infalling cloud does not reach the AGN. 

In the literature we find comparable galaxy groups in a similar dynamical state as NGC~5044, having a centrally peaked, relaxed X-ray morphology with a short central cooling time (\citealp{Hudson2010-bk,Voit2015-uc,OSullivan2017-zr}). However, examples like NGC~4636 and NGC~5846 show much less cold, molecular gas than NGC~5044. The radio power at GHz frequencies of NGC~5044 is about an order of magnitude higher than these two galaxies, which should indicate stronger feedback, preventing the formation of cold gas (\citealp{OSullivan2018-yk}). 
Also the cavity power and star formation rate are highest for NGC~5044, compared to NGC4636 and NGC5846. 
However, the cavity radio spectrum shows by far the steepest spectral index for NGC~5044 ($\alpha$ of 1.9 or higher, while the other two galaxies have $\sim 0.65$, see \citealp{OSullivan2011-xb}). 
This indicates that the visible signs of major feedback are not recent for NGC~5044, while the other two galaxies had more recent outbursts of the AGN. We note that other cavities in NGC~5044 reported by \cite{David2009-hn,David2017-ig} appear to be related to more recent AGN outbursts, but are much smaller then the one with the steep radio spectrum. 
NGC~5044 may thus have had longer to replenish its CO reservoir since the last major AGN outburst.

\section{Summary}\label{ch:summary}
We report the analysis of the ALMA 12m extended array data and the Atacama Compact Array (7m plus total power array) CO(2-1) data for the central dominant galaxy of the nearby galaxy group NGC~5044. \cite{David2014-jn} found molecular gas in this galaxy, accumulated in clouds, but the total emission detected in this high resolution interferometer observation was far below the single dish measurement. This discrepancy motivated the additional observation of NGC~5044 with ACA to cover short baselines. Our results can be summarized as follows:

\begin{itemize}
    \item ACA detects all of the emission that was missed by the 12m ALMA observation and the ACA measurement is consistent with the IRAM single dish measurements. The total molecular gas mass within a circular aperture of $\SI{15}{\arcsec}$ radius is $\SI{6.5(2)e7}{M_\odot}$.
    \item Extended emission beyond the IRAM single dish beam size of $\SI{6}{\arcsec}$ radius is detected in the SW-NE direction with relative velocities of $\SI{-100}{km\,s^{-1}}$.  The velocity range of the total CO(2-1) emission ranges from $-550$ to $\SI{230}{km\,s^{-1}}$.
    \item The \verb|CLUMPFIND| algorithm detects 10 individual clouds in the ACA data, which account for about 70\% of the total emission. 
    \item An absorption feature (reported in \citealp{David2014-jn}) can be quantified in great detail in the high angular resolution dataset. The emission from the continuum source (SMBH) features two absorption lines at 255 and $\SI{265}{km\,s^{-1}}$, where the latter one is strongest in the western region, and not detected in the eastern region. This means that the continuum emission region is only slightly smaller than the PSF.
    Assuming virial equilibrium, we estimate the  mass of the two absorbing clouds from their line widths as a few thousand to $\SI{e4}{M_\odot}$. The two clouds are consistent with being in pressure equilibrium with the surrounding gas.
    \item In projection, the absorbing structures lie within $\SI{5}{pc}$ (the continuum image resolution) of the central source. By combining the thermal evaporation timescale with constraints from the infall velocity, we argue that the distance, from the absorbing clouds to the AGN, is less than $\SI{19}{pc}$. Assuming a uniform distribution of clouds, there is a 80\% chance that the absorbing clouds are located within the sphere of influence of the SMBH.
\end{itemize}

\acknowledgments
This paper makes use of the following ALMA data: 2011.0.00735.S, 2016.1.00533.S, 2016.2.00134.S, 2017.1.00784.S. ALMA is a partnership of ESO (representing its member states), NSF (USA) and NINS (Japan), together with NRC (Canada), MOST and ASIAA (Taiwan), and KASI (Republic of Korea), in cooperation with the Republic of Chile. The Joint ALMA Observatory is operated by ESO, AUI/NRAO and NAOJ. The National Radio Astronomy Observatory is a facility of the National Science Foundation operated under cooperative agreement by Associated Universities, Inc.
Support for this work was provided by the National Aeronautics and Space Administration (NASA) through \textit{Chandra} award Nos. GO6-17121X, AR9-20013X, issued by the \textit{Chandra} X-ray Observatory Center (CXC), which is operated by the Smithsonian Astrophysical Observatory (SAO) for and on behalf of NASA under contract NAS8-03060.
This research made use of NASA's Astrophysical Data System Bibliographic Services and Astropy, a community-developed core Python package for astronomy.
Basic research in radio astronomy at the Naval Research Laboratory is supported by 6.1 Base funding. 
WF and CJ acknowledge support by the High Resolution Camera program of NASA contract NAS8-03060 and the Smithsonian Astrophysical Observatory. 
JL acknowledges support from the Research Grants Council of Hong Kong through grant 17304817 for the conduct of this work.
MS acknowledges support from the NSF grant 1714764.

\bibliographystyle{aasjournal}
\bibliography{Paperpile.bib}

\begin{thebibliography}{}
\expandafter\ifx\csname natexlab\endcsname\relax\def\natexlab#1{#1}\fi
\providecommand{\url}[1]{\href{#1}{#1}}
\providecommand{\dodoi}[1]{doi:~\href{http://doi.org/#1}{\nolinkurl{#1}}}
\providecommand{\doeprint}[1]{\href{http://ascl.net/#1}{\nolinkurl{http://ascl.net/#1}}}
\providecommand{\doarXiv}[1]{\href{https://arxiv.org/abs/#1}{\nolinkurl{https://arxiv.org/abs/#1}}}

\bibitem[{Bertoldi \& McKee(1992)}]{Bertoldi1992-zf}
Bertoldi, F., \& McKee, C.~F. 1992, Astrophys. J.

\bibitem[{Bolatto {et~al.}(2013)Bolatto, Wolfire, \& Leroy}]{Bolatto2013-oe}
Bolatto, A.~D., Wolfire, M., \& Leroy, A.~K. 2013, Annu. Rev. Astron.
  Astrophys., 51, 207

\bibitem[{Braine \& Combes(1992)}]{Braine1992-or}
Braine, J., \& Combes, F. 1992, Astron. Astrophys.

\bibitem[{Brighenti {et~al.}(2015)Brighenti, Mathews, \&
  Temi}]{Brighenti2015-cu}
Brighenti, F., Mathews, W.~G., \& Temi, P. 2015, ApJ, 802, 118

\bibitem[{Br{\"u}ggen \& Scannapieco(2016)}]{Bruggen2016-sh}
Br{\"u}ggen, M., \& Scannapieco, E. 2016, Astrophys. J., 822, 31

\bibitem[{Cornwell(2008)}]{Cornwell2008-no}
Cornwell, T.~J. 2008, IEEE J. Sel. Top. Signal Process., 2, 793

\bibitem[{Dame(2011)}]{Dame2011-hk}
Dame, T.~M. 2011.
\newblock \doarXiv{1101.1499}

\bibitem[{David {et~al.}(2009)David, Jones, Forman, Nulsen, Vrtilek,
  O'Sullivan, Giacintucci, \& Raychaudhury}]{David2009-hn}
David, L.~P., Jones, C., Forman, W., {et~al.} 2009, \apj, 705, 624

\bibitem[{David {et~al.}(2017)David, Vrtilek, O'Sullivan, Jones, Forman, \&
  Sun}]{David2017-ig}
David, L.~P., Vrtilek, J., O'Sullivan, E., {et~al.} 2017, ApJ, 842, 84

\bibitem[{David {et~al.}(2014)David, Lim, Forman, Vrtilek, Combes, Salome,
  Edge, Hamer, Jones, Sun, O'Sullivan, Gastaldello, Bardelli, Temi, Schmitt,
  Ohyama, Mathews, Brighenti, Giacintucci, \& Trung}]{David2014-jn}
David, L.~P., Lim, J., Forman, W., {et~al.} 2014, \apj, 792, 94

\bibitem[{Dobbs {et~al.}(2011)Dobbs, Burkert, \& Pringle}]{Dobbs2011-ds}
Dobbs, C.~L., Burkert, A., \& Pringle, J.~E. 2011, Mon. Not. R. Astron. Soc.,
  413, 2935

\bibitem[{Edge(2001)}]{Edge2001-mi}
Edge, A.~C. 2001, Mon. Not. R. Astron. Soc.

\bibitem[{Fabian(1994)}]{Fabian1994-oe}
Fabian, A.~C. 1994, \araa, 32, 277

\bibitem[{Field {et~al.}(2011)Field, Blackman, \& Keto}]{Field2011-kn}
Field, G.~B., Blackman, E.~G., \& Keto, E.~R. 2011, Mon. Not. R. Astron. Soc.,
  416, 710

\bibitem[{Gaspari {et~al.}(2013)Gaspari, Ruszkowski, \& Oh}]{Gaspari2013-eq}
Gaspari, M., Ruszkowski, M., \& Oh, S.~P. 2013, Mon. Not. R. Astron. Soc., 432,
  3401

\bibitem[{Gaspari {et~al.}(2017)Gaspari, Temi, \& Brighenti}]{Gaspari2017-ev}
Gaspari, M., Temi, P., \& Brighenti, F. 2017, Mon. Not. R. Astron. Soc., 466,
  677

\bibitem[{Gaspari {et~al.}(2018)Gaspari, McDonald, Hamer, Brighenti, Temi,
  Gendron-Marsolais, Hlavacek-Larrondo, Edge, Werner, Tozzi, Sun, Stone,
  Tremblay, Hogan, Eckert, Ettori, Yu, Biffi, \& Planelles}]{Gaspari2018-ni}
Gaspari, M., McDonald, M., Hamer, S.~L., {et~al.} 2018, ApJ, 854, 167

\bibitem[{Giacintucci {et~al.}(2011)Giacintucci, O'Sullivan, Vrtilek, David,
  Raychaudhury, Venturi, Athreya, Clarke, Murgia, Mazzotta, Gitti, Ponman,
  Ishwara-Chandra, Jones, \& Forman}]{Giacintucci2011-ru}
Giacintucci, S., O'Sullivan, E., Vrtilek, J., {et~al.} 2011, \apj, 732, 95

\bibitem[{Hudson {et~al.}(2010)Hudson, Mittal, Reiprich, Nulsen, Andernach, \&
  Sarazin}]{Hudson2010-bk}
Hudson, D.~S., Mittal, R., Reiprich, T.~H., {et~al.} 2010, {\aa}, 513, A37

\bibitem[{Lim {et~al.}(2008)Lim, Ao, \& {Dinh-V-Trung}}]{Lim2008-fa}
Lim, J., Ao, Y., \& {Dinh-V-Trung}. 2008, \apj, 672, 252

\bibitem[{Lim {et~al.}(2017)Lim, {Dinh-V-Trung}, Vrtilek, David, \&
  Forman}]{Lim2017-ig}
Lim, J., {Dinh-V-Trung}, Vrtilek, J., David, L.~P., \& Forman, W. 2017,
  Astrophys. J., 850, 31

\bibitem[{McMullin {et~al.}(2007)McMullin, Waters, Schiebel, Young, \&
  Golap}]{McMullin2007-ed}
McMullin, J.~P., Waters, B., Schiebel, D., Young, W., \& Golap, K. 2007, in
  Astronomical Society of the Pacific Conference Series, Vol. 376, Astronomical
  Data Analysis Software and Systems {XVI}, ed. R.~A. Shaw, F.~Hill, \& D.~J.
  Bell, 127

\bibitem[{McNamara \& Nulsen(2007)}]{McNamara2007-xa}
McNamara, B.~R., \& Nulsen, P. E.~J. 2007, Annu. Rev. Astron. Astrophys., 45,
  117

\bibitem[{McNamara {et~al.}(2016)McNamara, Russell, Nulsen, Hogan, Fabian,
  Pulido, \& Edge}]{McNamara2016-ux}
McNamara, B.~R., Russell, H.~R., Nulsen, P. E.~J., {et~al.} 2016, Astrophys.
  J., 830

\bibitem[{Olivares {et~al.}(2019)Olivares, Salome, Combes, Hamer, Guillard,
  Lehnert, Polles, Beckmann, Dubois, Donahue, Edge, Fabian, McNamara, Rose,
  Russell, Tremblay, Vantyghem, Canning, Ferland, Godard, Peirani, \& des
  Forets}]{Olivares2019-rs}
Olivares, V., Salome, P., Combes, F., {et~al.} 2019, Astron. Astrophys. Suppl.
  Ser., 631, A22

\bibitem[{O'Sullivan {et~al.}(2011)O'Sullivan, Giacintucci, David, Gitti,
  Vrtilek, Raychaudhury, \& Ponman}]{OSullivan2011-xb}
O'Sullivan, E., Giacintucci, S., David, L.~P., {et~al.} 2011, Astrophys. J.,
  735

\bibitem[{O'Sullivan {et~al.}(2017)O'Sullivan, Ponman, Kolokythas,
  Raychaudhury, Babul, Vrtilek, David, Giacintucci, Gitti, \&
  Haines}]{OSullivan2017-zr}
O'Sullivan, E., Ponman, T.~J., Kolokythas, K., {et~al.} 2017, Mon. Not. R.
  Astron. Soc., 472, 1482

\bibitem[{O'Sullivan {et~al.}(2018)O'Sullivan, Combes, Salom{\'e}, David,
  Babul, Vrtilek, Lim, Olivares, Raychaudhury, \&
  Schellenberger}]{OSullivan2018-yk}
O'Sullivan, E., Combes, F., Salom{\'e}, P., {et~al.} 2018, Astron. Astrophys.
  Suppl. Ser., 618, A126

\bibitem[{Pineda {et~al.}(2009)Pineda, Rosolowsky, \& Goodman}]{Pineda2009-ir}
Pineda, J.~E., Rosolowsky, E.~W., \& Goodman, A.~A. 2009, ApJ, 699, L134

\bibitem[{Rose {et~al.}(2019{\natexlab{a}})Rose, Edge, Combes, Gaspari, \&
  {others}}]{Rose2019-tu}
Rose, T., Edge, A.~C., Combes, F., Gaspari, M., \& {others}.
  2019{\natexlab{a}}, Mon. Not. R. Astron. Soc.

\bibitem[{Rose {et~al.}(2019{\natexlab{b}})Rose, Edge, Combes, Gaspari, Hamer,
  Nesvadba, Peck, Sarazin, Tremblay, Baum, Bremer, McNamara, O'Dea, Oonk,
  Russell, Salom{\'e}, Donahue, Fabian, Ferland, Mittal, \&
  Vantyghem}]{Rose2019-au}
Rose, T., Edge, A.~C., Combes, F., {et~al.} 2019{\natexlab{b}}, Mon. Not. R.
  Astron. Soc., 489, 349

\bibitem[{Russell {et~al.}(2014)Russell, McNamara, Edge, Nulsen, Main,
  Vantyghem, Combes, Fabian, Murray, Salom{\'e}, Wilman, Baum, Donahue, O'Dea,
  Oonk, Tremblay, \& Voit}]{Russell2014-su}
Russell, H.~R., McNamara, B.~R., Edge, A.~C., {et~al.} 2014, Astrophys. J.,
  784, 78

\bibitem[{Russell {et~al.}(2016)Russell, McNamara, Fabian, Nulsen, Edge,
  Combes, Murray, Parrish, Salom{\'e}, Sanders, \& {Others}}]{Russell2016-ek}
Russell, H.~R., McNamara, B.~R., Fabian, A.~C., {et~al.} 2016, Mon. Not. R.
  Astron. Soc., 458, 3134

\bibitem[{Russell {et~al.}(2017{\natexlab{a}})Russell, McDonald, McNamara,
  Fabian, Nulsen, Bayliss, Benson, Brodwin, Carlstrom, Edge, \&
  {Others}}]{Russell2017-xk}
Russell, H.~R., McDonald, M., McNamara, B.~R., {et~al.} 2017{\natexlab{a}},
  Astrophys. J., 836, 130

\bibitem[{Russell {et~al.}(2017{\natexlab{b}})Russell, McNamara, Fabian,
  Nulsen, Combes, Edge, Hogan, McDonald, Salom{\'e}, Tremblay, \&
  Vantyghem}]{Russell2017-zd}
Russell, H.~R., McNamara, B.~R., Fabian, A.~C., {et~al.} 2017{\natexlab{b}},
  Mon. Not. R. Astron. Soc., 472, 4024

\bibitem[{Salom{\'e} \& Combes(2003)}]{Salome2003-ge}
Salom{\'e}, P., \& Combes, F. 2003, Astron. Astrophys. Suppl. Ser., 412, 657

\bibitem[{Salom{\'e} {et~al.}(2008)Salom{\'e}, Combes, Revaz, Edge, Hatch,
  Fabian, \& Johnstone}]{Salome2008-kr}
Salom{\'e}, P., Combes, F., Revaz, Y., {et~al.} 2008, Astron. Astrophys. Suppl.
  Ser., 484, 317

\bibitem[{Sofia {et~al.}(2004)Sofia, Lauroesch, Meyer, \&
  Cartledge}]{Sofia2004-ck}
Sofia, U.~J., Lauroesch, J.~T., Meyer, D.~M., \& Cartledge, S. I.~B. 2004,
  Astrophys. J., 605, 272

\bibitem[{Solomon {et~al.}(1987)Solomon, Rivolo, \& {others}}]{Solomon1987-yg}
Solomon, P.~M., Rivolo, A.~R., \& {others}. 1987, Astrophys. J.

\bibitem[{Soto {et~al.}(2016)Soto, Lilly, Bacon, Richard, \&
  Conseil}]{Soto2016-ux}
Soto, K.~T., Lilly, S.~J., Bacon, R., Richard, J., \& Conseil, S. 2016, Mon.
  Not. R. Astron. Soc., 458, 3210

\bibitem[{Temi {et~al.}(2018)Temi, Amblard, Gitti, Brighenti, Gaspari, Mathews,
  \& David}]{Temi2018-rr}
Temi, P., Amblard, A., Gitti, M., {et~al.} 2018, ApJ, 858, 17

\bibitem[{Tonry {et~al.}(2001)Tonry, Dressler, Blakeslee, Ajhar, Fletcher,
  Luppino, Metzger, \& Moore}]{Tonry2001-wi}
Tonry, J.~L., Dressler, A., Blakeslee, J.~P., {et~al.} 2001, Astrophys. J.,
  546, 681

\bibitem[{Tremblay {et~al.}(2016)Tremblay, Oonk, Combes, Salom{\'e}, O'Dea,
  Baum, Voit, Donahue, McNamara, Davis, McDonald, Edge, Clarke,
  Galv{\'a}n-Madrid, Bremer, Edwards, Fabian, Hamer, Li, Maury, Russell,
  Quillen, Urry, Sanders, \& Wise}]{Tremblay2016-xg}
Tremblay, G.~R., Oonk, J. B.~R., Combes, F., {et~al.} 2016, Nature, 534, 218

\bibitem[{Vantyghem {et~al.}(2017)Vantyghem, McNamara, Edge, Combes, Russell,
  Fabian, Hogan, McDonald, Nulsen, \& Salom{\'e}}]{Vantyghem2017-ym}
Vantyghem, A.~N., McNamara, B.~R., Edge, A.~C., {et~al.} 2017, ApJ, 848, 101

\bibitem[{Voit {et~al.}(2015{\natexlab{a}})Voit, Bryan, O'Shea, \&
  Donahue}]{Voit2015-op}
Voit, G.~M., Bryan, G.~L., O'Shea, B.~W., \& Donahue, M. 2015{\natexlab{a}},
  Astrophys. J., 808

\bibitem[{Voit {et~al.}(2015{\natexlab{b}})Voit, Mark~Voit, Donahue, O'Shea,
  Bryan, Sun, \& Werner}]{Voit2015-uc}
Voit, G.~M., Mark~Voit, G., Donahue, M., {et~al.} 2015{\natexlab{b}},
  Astrophys. J., 803, L21

\bibitem[{Werner {et~al.}(2014)Werner, Oonk, Sun, Nulsen, Allen, Canning,
  Simionescu, Hoffer, Connor, Donahue, Edge, Fabian, von~der Linden, Reynolds,
  \& Ruszkowski}]{Werner2014-vw}
Werner, N., Oonk, J. B.~R., Sun, M., {et~al.} 2014, Mon. Not. R. Astron. Soc.,
  439, 2291

\bibitem[{Williams {et~al.}(1994)Williams, De~Geus, \& Blitz}]{Williams1994-uc}
Williams, J.~P., De~Geus, E.~J., \& Blitz, L. 1994, Astrophys. J.

\end{thebibliography}

\appendix
\section{CO cloud maps}

\begin{figure}[b]
\begin{center}
\includegraphics[width=0.99\linewidth]{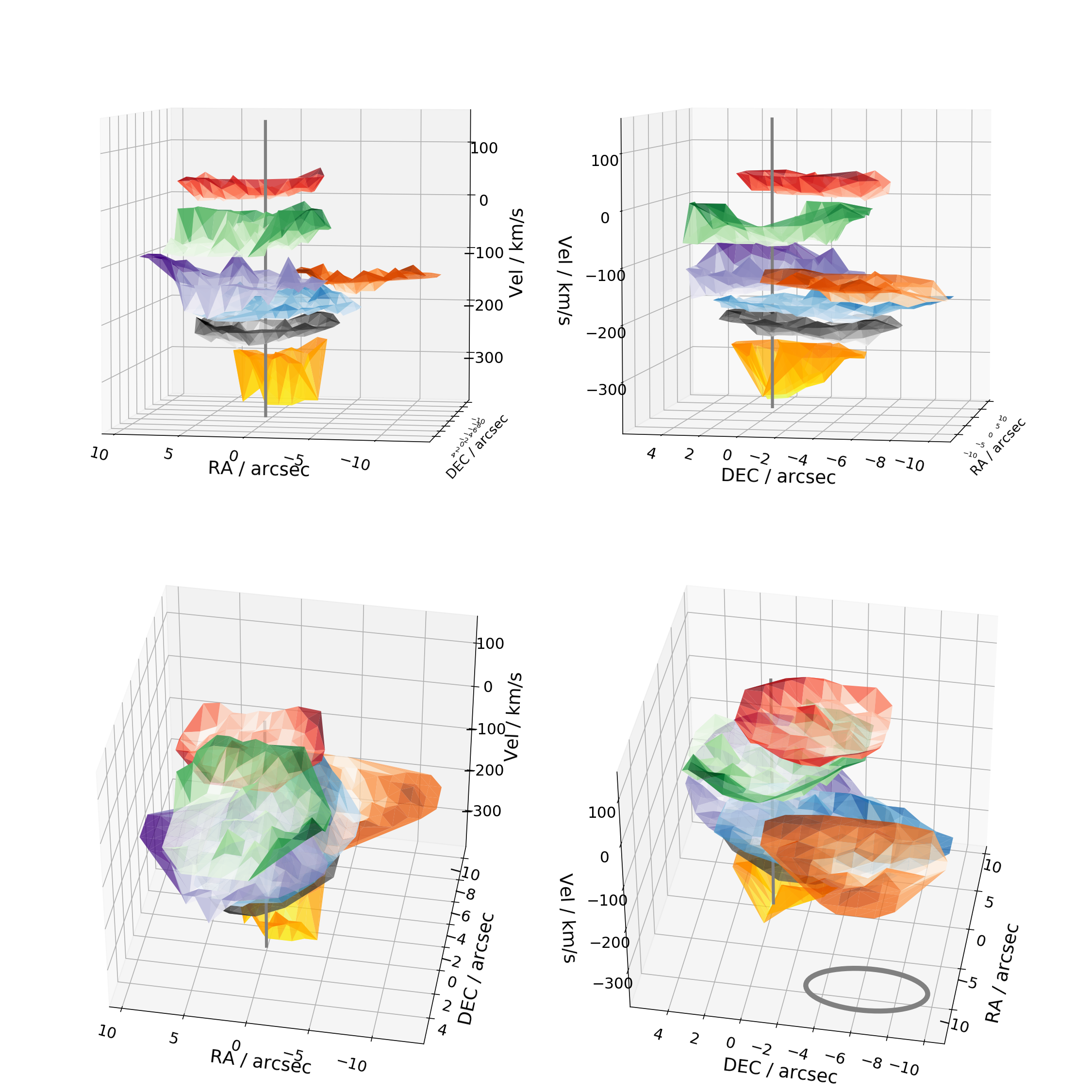}
\caption{3D representation of the CLUMPFIND detected clouds in the ACA data in spatial and velocity space for different angles. 
The clouds are spatially overlapping and can only be separated in velocity space.
Colors correspond to cloud IDs (defined in Tab. \ref{tab:clouds}): 1 -- red, 2 -- green, 3 -- blue, 4 -- purple, 5 -- grey, 6 -- orange, 7 -- yellow. The ellipse in the bottom right panel shows the ACA beam. \label{fig:cloud3d}}
\end{center}
\end{figure}

\begin{figure*}
\begin{center}
\includegraphics[width=0.99\textwidth]{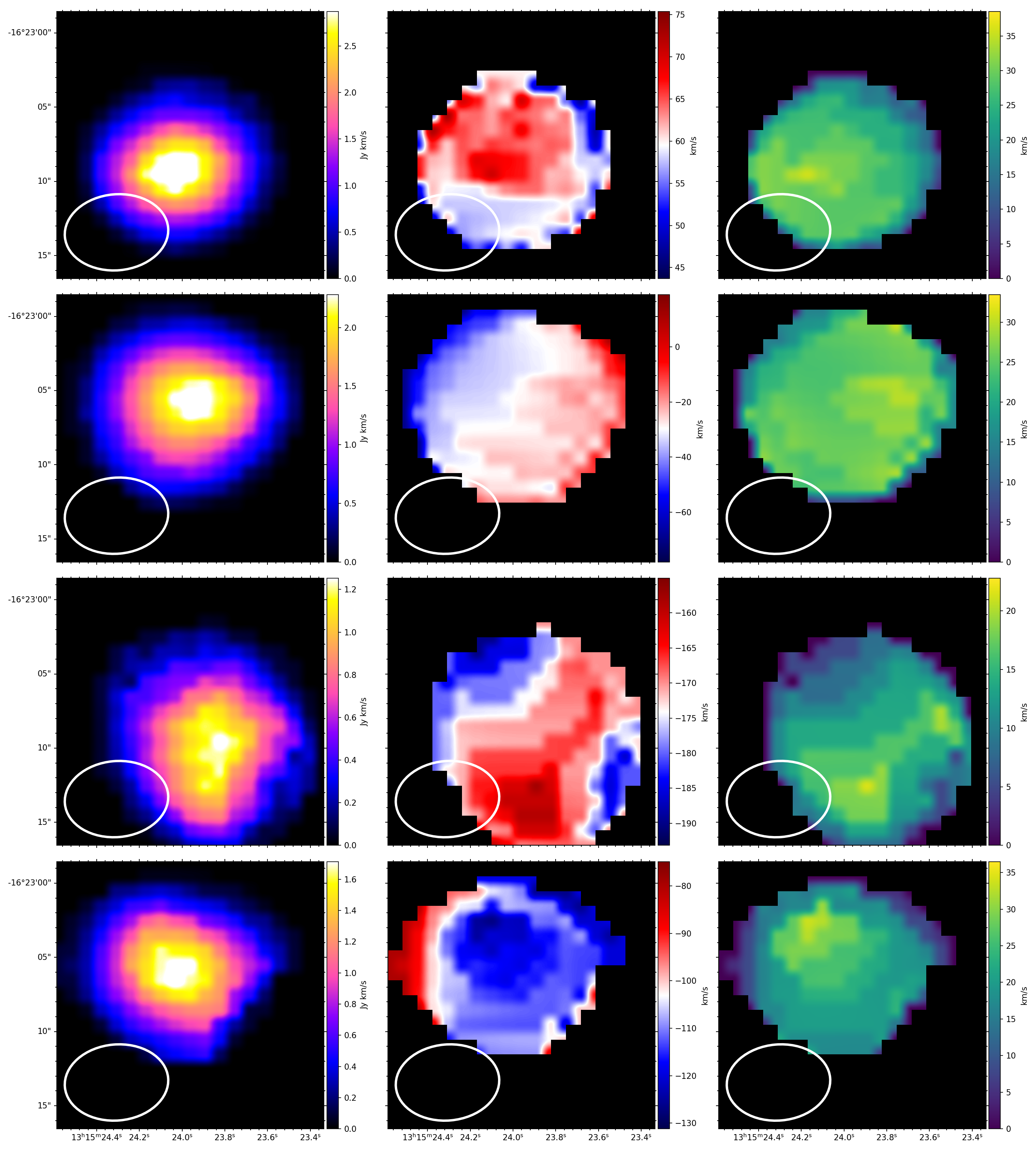}
\caption{\nth{0} (left), \nth{1} (middle), and \nth{2} (right) moment maps of the detected clouds. From top to bottom row, the cloud IDs are 1 through 4 (see Tab. \ref{tab:clouds}).  The beamsize is shown in the lower left corner. \label{fig:cloud_moments}}
\end{center}
\end{figure*}

\begin{figure*}
\begin{center}
\includegraphics[width=0.99\textwidth]{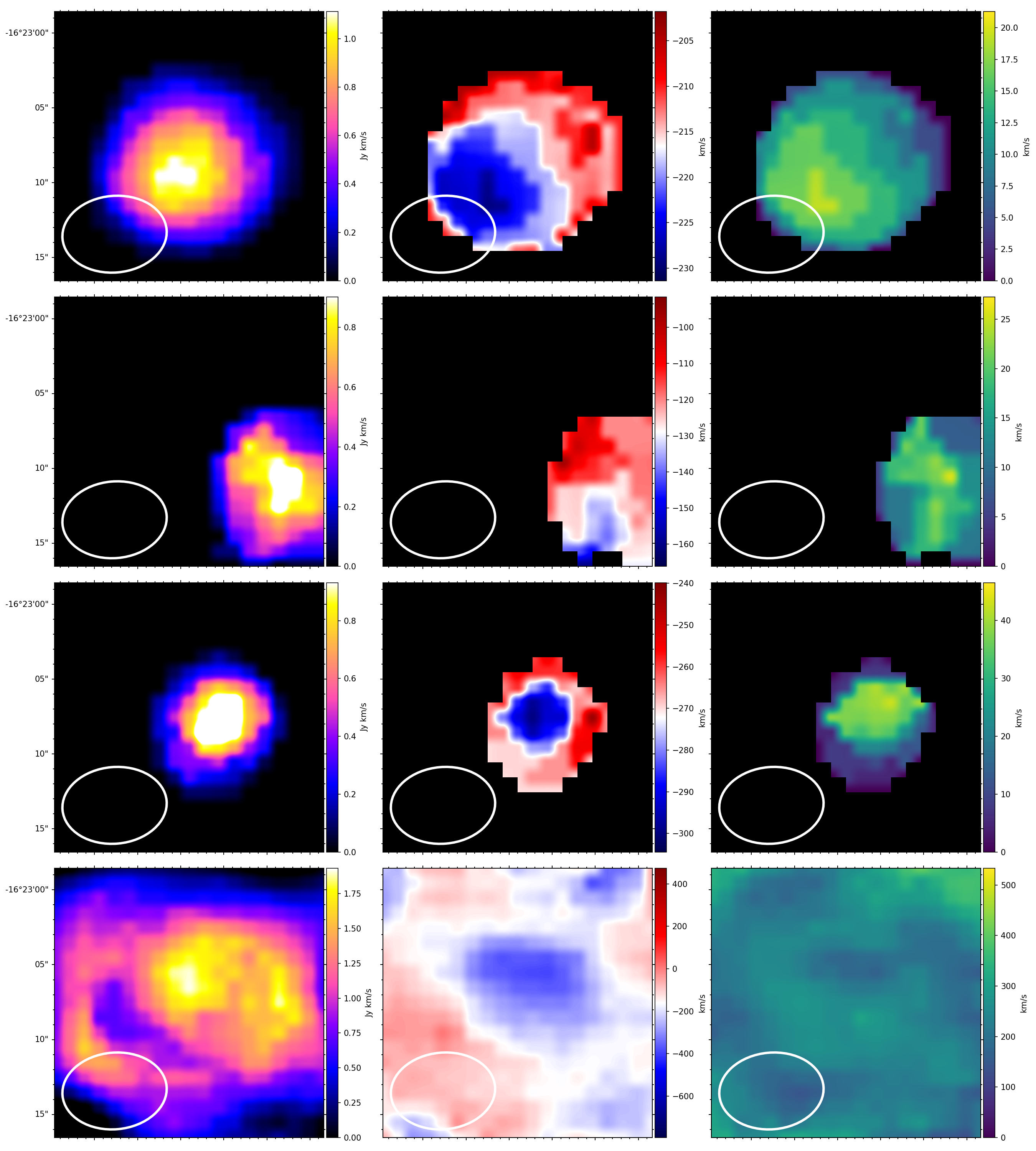}
\caption{\nth{0} (left), \nth{1} (middle), and \nth{2} (right) moment maps of the detected clouds. From top to bottom row, the cloud IDs are 5, 6, 7, and 0 (see Tab. \ref{tab:clouds}). \label{fig:cloud_moments2}}
\end{center}
\end{figure*}

\end{document}